\def\gsim{~\rlap{$>$}{\lower 1.0ex\hbox{$\sim$}}}
\def\lsim{~\rlap{$<$}{\lower 1.0ex\hbox{$\sim$}}}

\def\wpm2{W m$^{-2}$}

\documentclass[12pt, preprint]{aastex}

\usepackage{apjfonts}

\shorttitle{MARVELS-1b:  A Short-Period, Brown Dwarf Desert Candidate from the SDSS-III MARVELS Planet Search}
\shortauthors{Lee et al.}

\begin{document}


\title{
MARVELS-1b:  A Short-Period, Brown Dwarf Desert Candidate from the SDSS-III MARVELS Planet Search
 }


\author{Brian L. Lee\altaffilmark{1}, 
Jian Ge\altaffilmark{1}, 
Scott W. Fleming\altaffilmark{1}, 
Keivan G. Stassun\altaffilmark{2,3}, 
B. Scott Gaudi\altaffilmark{4},
Rory Barnes\altaffilmark{5}, 
Suvrath Mahadevan\altaffilmark{1,6,7},
Jason D. Eastman\altaffilmark{4}, 
Jason Wright\altaffilmark{6,7},
Robert J. Siverd\altaffilmark{4},
Bruce Gary\altaffilmark{2},
Luan Ghezzi\altaffilmark{8},
Chris Laws\altaffilmark{5},
John P. Wisniewski\altaffilmark{5},
G. F. Porto de Mello\altaffilmark{9},
Ricardo L. C. Ogando\altaffilmark{8},
Marcio A. G. Maia\altaffilmark{8},
Luiz Nicolaci da Costa\altaffilmark{8},
Thirupathi Sivarani\altaffilmark{1,10},
Joshua Pepper\altaffilmark{2},
Duy Cuong Nguyen\altaffilmark{1},
Leslie Hebb\altaffilmark{2},
Nathan De Lee\altaffilmark{1},
Ji Wang\altaffilmark{1},
Xiaoke Wan\altaffilmark{1},
Bo Zhao\altaffilmark{1},
Liang Chang\altaffilmark{1},
John Groot\altaffilmark{1},
Frank Varosi\altaffilmark{1},
Fred Hearty\altaffilmark{1},
Kevin Hanna\altaffilmark{1},
J. C. van Eyken\altaffilmark{11},
Stephen R. Kane\altaffilmark{11},
Eric Agol\altaffilmark{5},
Dmitry Bizyaev\altaffilmark{12},
John J. Bochanski\altaffilmark{13},
Howard Brewington\altaffilmark{12},
Zhiping Chen\altaffilmark{1},
Erin Costello\altaffilmark{1},
Liming Dou\altaffilmark{1},
Daniel J. Eisenstein\altaffilmark{14},
Adam Fletcher\altaffilmark{1},
Eric B. Ford\altaffilmark{1},
Pengcheng Guo\altaffilmark{1},
Jon A. Holtzman\altaffilmark{15},
Peng Jiang\altaffilmark{1},
R. French Leger\altaffilmark{5},
Jian Liu\altaffilmark{1},
Daniel C. Long\altaffilmark{12},
Elena Malanushenko\altaffilmark{12},
Viktor Malanushenko\altaffilmark{12},
Mohit Malik\altaffilmark{1},
Daniel Oravetz\altaffilmark{12},
Kaike Pan\altaffilmark{12},
Pais Rohan\altaffilmark{1},
Donald P. Schneider\altaffilmark{6,7},
Alaina Shelden\altaffilmark{12},
Stephanie A. Snedden\altaffilmark{12},
Audrey Simmons\altaffilmark{12},
B. A. Weaver\altaffilmark{16},
David H. Weinberg\altaffilmark{4},
Ji-Wei Xie\altaffilmark{1}}

\email{lee@astro.ufl.edu}
\altaffiltext{1}{Department of Astronomy, University of Florida, 211 Bryant Space Science Center, Gainesville, FL, 32611-2055, USA}
\altaffiltext{2}{Department of Physics and Astronomy, Vanderbilt University, Nashville, TN 37235, USA}
\altaffiltext{3}{Department of Physics, Fisk University, 1000 17th Ave. N., Nashville, TN 37208, USA}
\altaffiltext{4}{Department of Astronomy, The Ohio State University, 140 West 18th Avenue, Columbus, OH 43210, USA}
\altaffiltext{5}{Department of Astronomy, University of Washington, Box 351580, Seattle, WA 98195-1580, USA}
\altaffiltext{6}{Center for Exoplanets and Habitable Worlds, The Pennsylvania State University, University Park, PA 16802, USA}
\altaffiltext{7}{Department of Astronomy and Astrophysics, The Pennsylvania State University, 525 Davey Laboratory, University Park, PA 16802, USA}
\altaffiltext{8}{Observat\'{o}rio Nacional, Rua General Jos\'{e} Cristino, 77, 20921-400 S\~{a}o Crist\'{o}v\~{a}o, Rio de Janeiro, RJ, Brazil}
\altaffiltext{9}{Universidade Federal do Rio de Janeiro, Observat\'{o}rio do Valongo, Ladeira do Pedro Ant\^{o}nio, 43, CEP: 20080-090, Rio de Janeiro, RJ, Brazil}
\altaffiltext{10}{Indian Institute of Astrophysics, II Block, Koramangala, Bangalore 560 034, India}
\altaffiltext{11}{NASA Exoplanet Science Institute, Caltech, MS 100-22, 770 South Wilson Avenue, Pasadena, CA 91125, USA}
\altaffiltext{12}{Apache Point Observatory, P.O. Box 59, Sunspot, NM 88349-0059, USA}
\altaffiltext{13}{MIT Kavli Institute for Astrophysics \& Space Research, Cambridge, MA 02139, USA}
\altaffiltext{14}{Steward Observatory, University of Arizona, Tucson, AZ 85121, USA}
\altaffiltext{15}{Department of Astronomy, MSC 4500, New Mexico State University, P.O. Box 30001, Las Cruces, NM 88003, USA}
\altaffiltext{16}{Center for Cosmology and Particle Physics, New York University, New York, NY, USA}

\begin{abstract}
We present a new short-period brown dwarf candidate around the star TYC~1240-00945-1.  This candidate was discovered in the first year of the Multi-object APO Radial Velocity Exoplanets Large-area Survey (MARVELS), which is part of the third phase of the Sloan Digital Sky Survey (SDSS-III), and we designate the brown dwarf as MARVELS-1b.  MARVELS uses the technique of dispersed fixed-delay interferometery to simultaneously obtain radial velocity measurements for 60 objects per field using a single, custom-built instrument that is fiber fed from the SDSS 2.5-m telescope.  From our 20 radial velocity measurements spread over a $\sim \! 370$\,d time baseline, we derive a Keplerian orbital fit with semi-amplitude $K=2.533 \pm 0.025$\,km~s$^{-1}$, period $P=5.8953 \pm 0.0004$\,d, and eccentricity consistent with circular.  Independent follow-up radial velocity data confirm the orbit.  Adopting a mass of $1.37 \pm 0.11\,M_\odot$ for the slightly evolved F9 host star, we infer that the companion has a minimum mass of $28.0 \pm 1.5\,M_{Jup}$, a semimajor axis $0.071 \pm 0.002$\,AU assuming an edge-on orbit, and is probably tidally synchronized.  We find no evidence for coherent instrinsic variability of the host star at the period of the companion at levels greater than a few millimagnitudes.  The companion has an {\it a priori} transit probability of $\sim \! 14\%$.  Although we find no evidence for transits, we cannot definitively rule them out for companion radii $\la \! 1\,R_{Jup}$.
\end{abstract}

\newpage

\section{INTRODUCTION} 

One of the first results to emerge from high-precision radial velocity (RV) surveys seeking substellar companions was the existence of a brown dwarf (BD) desert:  a paucity of close ($a\la \! 5$\,AU) brown dwarf ($13\,M_{Jup}\!\la\!M \la 80\,M_{Jup}$) companions to solar-type stars, relative to more common stellar mass companions \citep{marcy00}.  Indeed, since they induce reflex radial velocity semiamplitudes of many hundreds of meters per second, such brown dwarf companions have been within the detection capabilities of these surveys for over two decades (e.g., \citealt{Campbell88}), yet to date only a few dozen are known (\citealt{reid08}).  On the other hand, as instrumentation has subsequently improved, first Jovian, and now terrestrial planetary companions in similar orbits have been found in relative abundance \citep{cumming08,mayor08,mayor09}.  The brown dwarf mass regime represents an apparent minimum in the mass distribution of close companions to solar-type stars.

Planetary companions are believed to form in circumstellar protoplanetary disks, whereas stellar companions are believed to form by concurrent collapse or fragmentation, so the brown dwarf desert is commonly interpreted as the gap between the largest mass objects that can be formed in disks, and the smallest mass clump that can collapse and/or fragment in the vicinity of a protostar.  Such a gap was by no means guaranteed to exist, and is perhaps surprising.  For example, numerous isolated BDs in star-forming regions have been found to possess protoplanetary disks, akin to the disks of young stars, suggesting that BDs form much as stars do \citep[e.g., ][]{Caballero07,Luhman08,Scholz08}. More generally, the mass function of isolated substellar objects in the field and clusters appears to be roughly flat in $\log{M}$ for masses down to at least $\sim\!20\,M_{Jup}$ \citep{luhman00,chabrier02}, whereas it is not clear what sets the upper limit for objects formed in protoplanetary disks \citep[e.g., ][]{boss01,idalin04,rafikov05,dodson-robinson09,kratter10}.

As such, details of the demographics of companions in the brown dwarf desert, including the aridity of the desert, the shape of the high-mass tail of the planetary companion mass function and the low-mass tail of the stellar companion mass function, as well as how these properties change with semimajor axis and primary mass, encode a wealth of information about the poorly understood physics of star and planet formation.  Additional processes such as tidal evolution and disk-planet migration can also affect these properties \citep[e.g., ][]{armitage02,matzner05}, thus can be investigated via brown dwarf desert statistics.

Unfortunately, despite its potential diagnostic power, and after more than twenty
years of precision radial velocity surveys, very little is known about
the brown dwarf desert, precisely because brown dwarf companions are
rare and so few such companions are known.  The California \& Carnegie
Planet Search finds an occurrence rate of 0.7\% $\pm$ 0.2\% from their
sample of $\sim \! 1000$ target stars (Vogt et al. 2002\nocite{Vogt02},
Patel et al. 2007\nocite{Patel07}), and the McDonald Observatory
Planet Search agrees, with a rate of 0.8\% $\pm$ 0.6\% from a search
sample of 250 stars \citep{Wittenmyer09}.  \citet{Gizis01} suggest
that brown dwarfs might not be as rare at wide separations \citep[see also][]{Metchev04}, although \citet{mccarthy04} find
a low rate of occurrence that is similar to that found for close
separations.  By extrapolating the mass functions of planets (on the low mass side) and stellar companions (on the high mass side) into the brown dwarf mass
regime, \citet{Grether06} find a mass of minimum occurrence (the
driest part of the brown dwarf desert) at $31^{+25}_{-18}\,M_{Jup}$.  They
further suggest that the location of this minimum may scale with host
star mass.  For instance, the only known BD eclipsing binary is a ``desert dweller'', consisting of a $\sim \! 60 \, M_{Jup}$ BD with a $\sim \! 35 \, M_{Jup}$ BD companion at a separation of $0.04$\,AU \citep{stassun06,stassun07}. 

To make further progress on understanding the properties of the brown dwarf desert, a much larger sample of brown dwarf companions is needed.  Furthermore, this larger sample must be drawn from a relatively uniform survey with a well-defined and homogeneous sample of primary target stars, so that the demographic properties of these companions can be reliably inferred.  Given the occurrence rate of $\sim \! 1\%$, a survey of $\sim \! 10000$ stars is needed to detect of order 100 brown dwarf companions.  Such an extensive survey would require a prohibitive amount of observing time with traditional echelle-based precision RV instruments, which can only target one object at a time.  Furthermore, in many cases the RV precisions that can be achieved with these instruments are far better than are needed to detect brown dwarf companions, implying that this is not the most efficient application of these instruments.

The Multi-object APO Radial Velocity Exoplanets Large-area Survey \citep[MARVELS;  ][]{Ge08} is a radial velocity survey of $\sim \! 11000$ stars ($\sim \! 10000$ dwarfs and subgiants, plus $\sim \! 1000$ giants) with $7.6 \! < \! V \! < \! 12$ over time baselines of $\sim \! 1.5$ years, with a stated goal of $< \! 30$\,m~s$^{-1}$ precision for the faintest stars.  It operates as one of the bright-time survey components of the Sloan Digital Sky Survey (SDSS) III, following on the legacy of the original SDSS \citep{york00}.  MARVELS uses the innovative instrumental technique of a dispersed fixed-delay interferometer \citep[DFDI;  see, e.g.,][]{erskine00,Ge02,Geetal02,vaneyken10} in order to simultaneously observe 60 objects at a time over a three degree field of view with a single instrument that is fiber fed from the SDSS 2.5-m Telescope \citep{gunn06} at Apache Point Observatory.  The fibers are fed through an interferometer, and both interferometer output beams are sent through a spectrograph with a resolving power $R \! \sim \! 12000$, producing fringing spectra over the wavelength range $\sim \! 500-570$\,nm.  Radial velocity information is imprinted in the phases of the fringes perpendicular to the dispersion axis of the spectrum due to a fixed variation in the interferometer delay along this direction.

By virtue of the large number of target stars, as well as uniform selection criteria described below, MARVELS is well suited to probe for rare companions.  MARVELS commenced operations with SDSS-III in Sep. 2008, and as of the end of the first year's data collection in Aug. 2009, had observed 780 stars with RV time series of more than 15 points.  In this paper, we report the first MARVELS brown dwarf candidate, which we designate MARVELS-1b, detected in orbit around the star TYC~1240-00945-1 (Tycho-2 star catalogue;  \citealt{hog00}).

\section{OVERVIEW OF SDSS-III MARVELS TARGET SELECTION}
\label{sec:targetsel}

The overall scope of MARVELS will be described in detail in future papers;  we present a brief outline here in order to provide the context for the field and target selection of the brown dwarf candidate.  MARVELS has been designed with an RV precision goal of $< \! 30$\,m~s$^{-1}$ in order to be able to discover a sample of $\sim \! 150$ new exoplanets, within a homogeneous parent sample of searched stars. By choosing a sample of target stars using a limited number of well-defined selection criteria, our sample suffers from minimal and well-understood biases, and can  increase the size of the largest statistically homogeneous exoplanet sample by a factor of a few over that currently available.  

MARVELS will run for six years during SDSS-III bright time, in a series of three cycles of self-contained two-year surveys.  Each cycle will have a similar stellar target selection strategy, designed to give good survey coverage of FGK dwarfs and similar parent samples in each two-year cycle, although with different target fields.  While in general this means only companions with up to $\sim \! 1.5$ year periods will be detected, the advantage of this strategy is that we need not wait the full six years to gather enough epochs per star to detect companions.  Also, this approach provides the opportunity to do major instrument upgrades at the end of each two-year cycle, without destroying the continuity of our RV measurements.

In order to collect enough photons to achieve $< \! 30$\,m~s$^{-1}$ statistical RV precision, the stars we monitor must in general be brighter than $V \! \sim \! 12$, although the precision at a given magnitude depends somewhat on stellar parameters as well.  For the 60 object multiplexing capability of the instrument during the first two years (Sep. 2008--Sep. 2010), we found most fields on the sky were sufficiently rich to fill all the object fibers, so half of our fields were selected to include a reference star of $8 \! < \! V \! < \! 12$ with a known RV signal (stable or planet-hosting).  By recovering the RV of the reference star, we can verify that the instrument is sufficiently stable to detect planetary companions.  To ensure survey observability across all right ascensions, the remaining fields were selected from areas with no reference stars.  Finally, we also selected some fields in the Kepler survey footprint \citep{borucki97} in order to have the potential to leverage the exquisite Kepler photometry for any stars targeted by both surveys.

In each individual field, we used the intersection of the GSC2.3 \citep{spagna06} and 2MASS \citep{skrutskie06} catalogs as our initial targets database, but because many of our fields are in the Galactic plane and contaminated by giants, we conducted a preselection program to identify and reject giants from the sample prior to beginning RV monitoring.  First, we performed a rough cut in magnitude and color, accepting only stars with $9.0 \! < \! V \! < \! 13.0$ and $(J-K) \! > \! 0.29$.  The faint magnitude limit rejects stars too faint for the survey, and the bright magnitude limit keeps the dynamic range small enough to avoid saturating the preselection observations.  The color cut eliminates most hot stars from consideration, since we cannot obtain sufficient RV precision to detect planetary companions on any star hotter than mid-F.  Second, we took spectral classification snapshots of the potential target stars using the SDSS double spectrographs \citep{uomoto99} mounted on the SDSS 2.5-m Telescope, which have $R \! \sim \! 1800$ and cover the wavelength range $390 \! < \! \lambda \! < \! 910$\,nm.

The preselection observations were processed using the SDSS two-dimensional and one-dimensional spectroscopic pipelines \citep{stoughton02}. The spectroscopic parameters $T_{\rm eff}$, $\log g$, and [Fe/H] were derived using the SEGUE Stellar Parameter Pipeline (SSPP;  \citealt{yslee08}). Each spectrum was manually inspected to validate the parameters and to identify obvious binaries and emission line objects.

The final 60 targets for each field were selected using the following method.  First, we only consider stars with $T_{\rm eff}<6250$\,K.  We dedicated 6 of the targets for observing giants, and identified the brightest available dwarfs and subgiants from $7.6 \! < \! V \! < \! 13.0$ to fill the other 54 targets, where dwarfs and subgiants are defined as having $\log{g} \! > \! 3.0$.  For $7.6 \! < \! V \! < \! 9.0$, we selected the targets for observation by conducting a literature search to reject known variable stars, and used a reduced proper motion (RPM) diagram to classify them as giants or dwarfs.  While we prefer to pick bright dwarfs, in practice this bright magnitude range is dominated by giants, and therefore the MARVELS giant sample is typically drawn from the bright magnitude bin.  For $9.0 \! < \! V \! < \! 13.0$, we ranked the stars by $V$ magnitude then picked the $\sim \! 54$ highest-ranked stars, although to avoid the survey being dominated by F-stars, we cap the number of stars with $5800$\,K$\!<\!T_{\rm eff}\!<\!6250$\,K at no more than 24 out of 60.  In practice, this combination of criteria usually completes our 60 target selection without going fainter than $V \! \sim \! 11.5-12.0$.  We do not impose selections based on the ages, activity levels, or metallicities of the stars.

We have recently learned that the original version of the SSPP code that we used for our target selection tends to overestimate $\log g$, particularly for cool temperatures of $T_{\rm eff} \! \la \! 5000$\,K.  While we are working on improved methods to better discriminate between dwarfs and giants for targeting in future survey cycles, our target sample for the first two year survey cycle is likely to have more giants than we desired;  we estimate that up to 30\% of targets in this sample could be giants due to the bias in the SSPP results.  Note that we do not use the primary properties of TYC~1240-00945-1 derived from the SSPP in our subsequent analysis;  we rely on the more accurate determinations from the detailed analysis of high-resolution spectra as described in \S\ref{sec:params}.  We only describe the SSPP target selection method here because our MARVELS targets for years 1 and 2 (including TYC~1240-00945-1) have been selected based on the SSPP results.

\section{OBSERVATIONS AND PROCESSING} 

\subsection{Primary Survey Observations with SDSS}
\label{sec:sdssobs} 

TYC~1240-00945-1 was part of the first two-year cycle of the SDSS-III MARVELS planet search program described above.  This target was selected for radial velocity monitoring using the preselection methodology and instrumentation described in \S\ref{sec:targetsel}.  In preselection observations for this star's field taken on Sep. 19, 2008, we obtained a series of five 7\,s and five 12\,s exposures of the target field, plus flat and arc lamp calibration exposures before and after this series.  From preselection, the star appeared to be a late F-dwarf (but see further details in \S\ref{sec:params}, which suggest it is starting to evolve into a subgiant) suitable for inclusion in the MARVELS RV monitoring.

Our discovery radial velocity observations were taken using the SDSS 2.5-m Telescope at Apache Point Observatory coupled to the MARVELS instrument, a 60 object fiber-fed DFDI \citep{Ge09}.  Our two-output interferometer produces two fringing spectra (``beams'') per object, over wavelengths $\sim \! 500-570$\,nm, with resolving power $R \! \sim \! 12000$.  The instrument is environmentally stabilized such that no iodine cell is needed in the stellar beam path, and instrument drift calibrations are simply taken before and after each stellar exposure.  TYC~1240-00945-1 was observed at 20 epochs from Nov. 7, 2008 to Nov. 11, 2009, as listed in Table \ref{tab:obsjournal}.  Exposures were 50 min., yielding an average of 500 photons per CCD pixel on each $4 \rm{k} \times 20$ pixel fringing spectrum.  The RV signal on TYC~1240-00945-1 was easily detected by eye in the RV curves from the first year of MARVELS.

MARVELS RVs are differential measurements, based on the shift of the fringing spectrum relative to a template epoch.  
The RVs were derived from our 20 fringing spectrum observations using the preliminary version of our MARVELS DFDI pipeline, which is based on software from earlier DFDI prototype instruments (e.g., Ge et al. 2006\nocite{Ge2006}).  We provide here a brief outline of the mechanics of the MARVELS-specific pipeline, but leave a full description to future techniques papers to be written on the overall performance of the MARVELS hardware and survey.

After performing standard multi-object spectroscopic preprocessing on each frame such as bias subtraction, flatfielding, and trimming out individual spectra, we proceed to straighten slanted spectral lines, straighten tilted traces, and divide out uneven slit illumination to produce clean images ready for analysis.  To remove a faint pattern of background fringes caused by the interferometer, we apply a low-pass filter, which leaves the fringes on stellar lines visible.  The pipeline seeks to measure the epoch-to-epoch shift in the two-dimensional fringing spectrum (i.e., a spectrum with sinusoidal modulations along the slit direction).  The shift induced by a stellar radial velocity change comprises two orthogonal components.  The first component, a small shift of the stellar absorption lines along the wavelength axis, is the shift that conventional Doppler planet search instrumentation seeks to measure.  The second component, a shift of the fringes on each absorption line along the spectrograph slit axis, is linearly proportional to the shift in the wavelength axis, but is amplified to a factor of a few times larger, and therefore provides most of the statistical leverage in our velocity measurement.  At any given wavelength, the fringe shift is related to the radial velocity by a multiplicative factor derived from measurements of the interferometer in the lab before commissioning the instrument.

We use $\chi^2$ minimization to determine the best-fit velocity shift for each epoch, relative to a template spectrum chosen to be the brightest one from the epochs that were observed.  Specifically, we determine the best-fit velocity shift that minimizes the shift of the spectrum along the wavelength and slit axes, relative to the template spectrum.  We also account for the barycentric correction during the RV extraction routine, ensuring that the $\chi^2$ minimizer does not need to search as far in velocity space as it would if the Earth's motion were not removed.   Wavelength and slit axis shifts between exposures induced by the instrument drift were measured from fringing spectra of a stable calibration source (a tungsten lamp shining through a temperature-stabilized I$_{\rm{2}}$ gas cell) taken before and after each stellar exposure, and the RV corrections due to these shifts are subtracted from each spectrum.  Because the epoch for the instrument drift RV zero-point differs from the epoch for the stellar RV zero-point, none of these differential RVs will have a value of exactly zero.

Because the interferometer splits the beam of each star, we record two separate spectra of each star on the CCD, and measure the RV from each of these spectra independently.  We shall differentiate between these two simultaneously observed RV curves by using the labels ``beam1'' and ``beam2.''  Although not all of the potential sources of systematic error would cause differences between the two beams' RV curves, comparison of the two beams does provide a partial consistency check of the quality of the data and the reduction pipeline.

\subsection{Photometric observations}

\label{sec:obsphotometry}

In order to check for intrinsic photometric variability indicating activity, as well as search for transits of the companion, we extracted the photometric time series data of  TYC~1240-00945-1 obtained by the Kilodegree Extremely Little Telescope (KELT) North transit survey \citep{Pepper07,siverd09}.  KELT consists of a $42$\,mm lens imaging a $26^{\circ} \times 26^{\circ}$ field of view onto a $4 \rm{k} \times 4\rm{k}$ CCD.  KELT uses a red-pass filter with a 50\% transmission point at $490$\,nm, which, when folded with the CCD response, yields an effective bandpass similar to $R$, but broader.  
 
The KELT data were processed as described in detail in \citet{fleming10}.  Briefly, after flat-fielding, relative photometry was extracted using the ISIS image subtraction package \citep{ISIS}, combined with point-spread fitting photometry using DAOPHOT \citep{daophot}.  We reduced the level of systematics present in the light curve by applying the Trend Filtering Algorithm (TFA;  \citealt{tfa}).  A few additional outlying measurements were removed before and after application of TFA.  Raw uncertainties on the individual points were scaled to force an ensemble of stars near the target to have a modal $\chi^2$/dof of unity for a constant fit.  
As in \citet{fleming10}, the target's $\chi^2$/dof was still not unity after this adjustment based on the ensemble, so we further scaled the target's error bars by a small amount ($\sim \! 10\%$) to force $\chi^2$/dof=1.  The final KELT light curve has 5036 data points taken between Nov. 15, 2006 and Jan. 17, 2010, with typical relative photometric precision of $\sim \! 1\%$.  

The Tycho catalog magnitudes \citep{hog00} of our targets are generally unreliable at $V \! > \! 11$.  In particular, we find the error bars can sometimes be underestimated at the level of several tenths of a magnitude, resulting in colors that do not agree with spectroscopically-determined values of $T_{\rm eff}$.  Therefore, we obtained absolute photometry to supersede and supplement the catalog colors.  
TYC~1240-00945-1 was observed in $BV$ under photometric conditions by the privately-owned Hereford Arizona Observatory (HAO) 11-inch on Jul. 29 and 31, 2009, together with a program of standards from \citet{landolt92}.  It was observed again by this telescope in $g'r'i'$ under photometric conditions on Jan. 12, 2010, together with a program of Landolt standards that had $u'g'r'i'z'$ calibrations from \citet{smith02}.  
This telescope is equipped with a $1.5\rm{k} \times 1\rm{k}$ CCD with a plate scale of 0.81" per pixel.  
For each program night, the standard star instrumental magnitudes were fit with a generic photometric equation (see \citealt{gary2010} for more information on calibration procedures at HAO), and the resulting fit used to calculate the apparent magnitudes of TYC~1240-00945-1;  typical standard star residuals relative to the fit were 0.01-0.02 magnitudes.  The resulting calibrated $BVg'r'i'$ are provided in Table \ref{tab:stellarparams}.  Magnitudes in $R_cI_c$ were estimated from the measured $g'r'i'$ by using the transformation equations tabulated in \citet{smith02};  the $R_cI_c$ estimates are also provided in Table \ref{tab:stellarparams}.

\subsection{Spectral Classification of Host Star}

In pursuit of a more detailed spectral classification of our candidate than is possible from our low-resolution SDSS spectrograph preselection observations, optical ($\sim \! 3600-10000$\,\AA) spectra of TYC~1240-00945-1 were obtained on Nov. 2, 2009 with the Apache Point Observatory 3.5-m telescope and ARC Echelle Spectrograph (ARCES;  \citealt{Wang03}).  We used the default $1\farcs6 \times 3\farcs2$ slit to obtain two moderate resolution ($R \! \sim \!31500$) spectra with signal-to-noise ratio (S/N) of $\sim \! 160$ per 1-D extracted pixel at $6500$\,\AA.
We extracted our APO classification spectra to 1-D using standard IRAF techniques and wavelength calibrated using ThAr lamp exposures
obtained immediately after each science exposure.  

We also used the high resolution ($R \! = \! 48000$) spectrograph FEROS \citep{Kaufer99} mounted at the MPG/ESO 2.2-m telescope in La Silla to obtain spectra of TYC~1240-00945-1. Two spectra, exposed for $3600$\,s and $4200$\,s respectively, were obtained in the wavelength interval $3500-9000$\,\AA, yielding a S/N $\sim \! 340$ per 1-D extracted pixel around $6600$\,\AA. 
These spectra were analyzed using the online FEROS Data Reduction System (DRS) and the standard calibration plan, where bias, flat-field and wavelength calibration lamp frames are observed in the afternoon.  \citet{Peloso05} checked the performance of the DRS by comparing the equivalent widths derived from solar spectra (observations of reflected sunlight from Ganymede) with those from the Solar Flux Atlas \citep{Kurucz84}.  They found that the two sets of  measurements are strongly correlated, with a correlation coefficient of $\rm{R} = 0.994$ and a standard deviation of $2.9$\,m\AA. The FEROS pipeline equivalent widths may thus be regarded as very robust. Furthermore, as the wavelength shift between the two observed spectra was found to be negligible ($13.1$\,m~s$^{-1}$), the two spectra were simply combined and shifted to the rest wavelength.

\subsection{Radial Velocity Follow-up}
\label{sec:rvfollowup}

To confirm this first substellar companion from MARVELS, as well as ascertain the quality of the radial velocities obtained with the MARVELS instrument relative to those measured using conventional echelle spectrograph technology, we used the High Resolution Spectrograph (HRS;  \citealt{Tull1998}) mounted on the 9-m Hobby-Eberly Telescope (HET;  \citealt{ramsey98}) to obtain additional precision RV measurements of TYC~1240-00945-1.  The candidate was observed in queue-scheduled mode \citep{shetrone07} with a Director's Discretionary Time allocation especially for this candidate, allowing for high-priority confirmation using just a few short ($\sim \! 15$-min.) exposures spread over several nights.  Nine measurements were taken in Dec.\ 2009 using an iodine cell for wavelength calibration, as well as one iodine-free template observation.  All spectra were taken with the $316$\,lines~mm$^{-1}$ grating with a central wavelength of $593.6$\,nm, leading to a resolving power $R \! \sim \! 60000$ and wavelength coverage $409 \! < \! \lambda \! < \! 782$\,nm.
Differential RVs were extracted from the HET spectra using a preliminary version of a new precise Doppler reduction pipeline (kindly provided by Debra Fischer) based on the principles outlined in \citet{butler96}.  This version of the pipeline was not yet optimized for the HRS fiber-fed spectrograph, and in particular used an instrumental profile description more appropriate for the slit-fed Hamilton spectrograph at Lick Observatory.  As a result, systematic errors in the radial velocities presented here are high, and do not reflect the full capabilities of either the iodine technique or the HRS.  The final measured radial velocities are given in Table \ref{tab:obsjournalhet}.

In addition, absolute radial velocities were obtained from the SMARTS 1.5-m telescope at CTIO.  The target was observed 9 times from Aug.--Dec. 2009 using the echelle spectrograph with no iodine cell, yielding $R \! \sim \! 42000$ and wavelength coverage $402 \! < \! \lambda \! < \! 730$\,nm.  Each observation spanned 30 minutes of total exposure time, subdivided into three 10-min.~exposures for cosmic ray removal.  RVs were extracted using an IDL based pipeline written by F.\ Walter and adapted by K.\ Stassun. The individual exposures were bias-subtracted, flat-field corrected using quartz lamp flats, and wavelength calibrated using ThAr lamp exposures bracketing the science exposures.  Typically, 35 good echelle orders spanning  $4800-7100$\,\AA, with a resolving power $R \! \sim \! 42000$, were extracted from each observation, with a typical S/N $\sim \! 30$ per resolution element.  Absolute RVs were measured via cross-correlation against an early-K giant radial velocity standard star, HD~223807, selected from the catalog of \citet{nidever02}, which was observed with the same instrument with S/N $\sim \! 100$. For each observation of TYC~1240-00945-1, cross correlation was performed order by order against the template, and the resulting 35 RV measurements from the individual orders were subjected to a sigma clipping based on the median absolute deviation.  After clipping, we typically were left with RV measurements from 20--25 orders, which were averaged for the final RV measurement at that epoch.  The measured absolute RVs are given in Table \ref{tab:obsjournalsmarts}.  We also applied this procedure to determine the RVs for six observations of the RV standard star obtained over the same time period;  we found the root mean square (RMS) scatter in the standard star's RV measurements was $70$\,m~s$^{-1}$, which we take as the current precision limit of RVs obtained with the SMARTS 1.5-m echelle, without the Iodine cell and with the current preliminary pipeline.  Note that, although the radial velocity standard is of a different spectral type than TYC~1240-00945-1, we expect that the systematic error that this mismatch produces will manifest itself primarily as an offset of $\sim \! 1-2$\,km~s$^{-1}$ added to all the absolute RV measurements, with a much lesser effect on the values of the RVs relative to each other.

\section{RADIAL VELOCITY ANALYSIS AND KEPLERIAN ORBITAL SOLUTION} 
 \label{sec:rvmeas} 

\subsection{MARVELS RADIAL VELOCITY DATA}
 \label{sec:marvelsrv}

\begin{figure}
\plotone{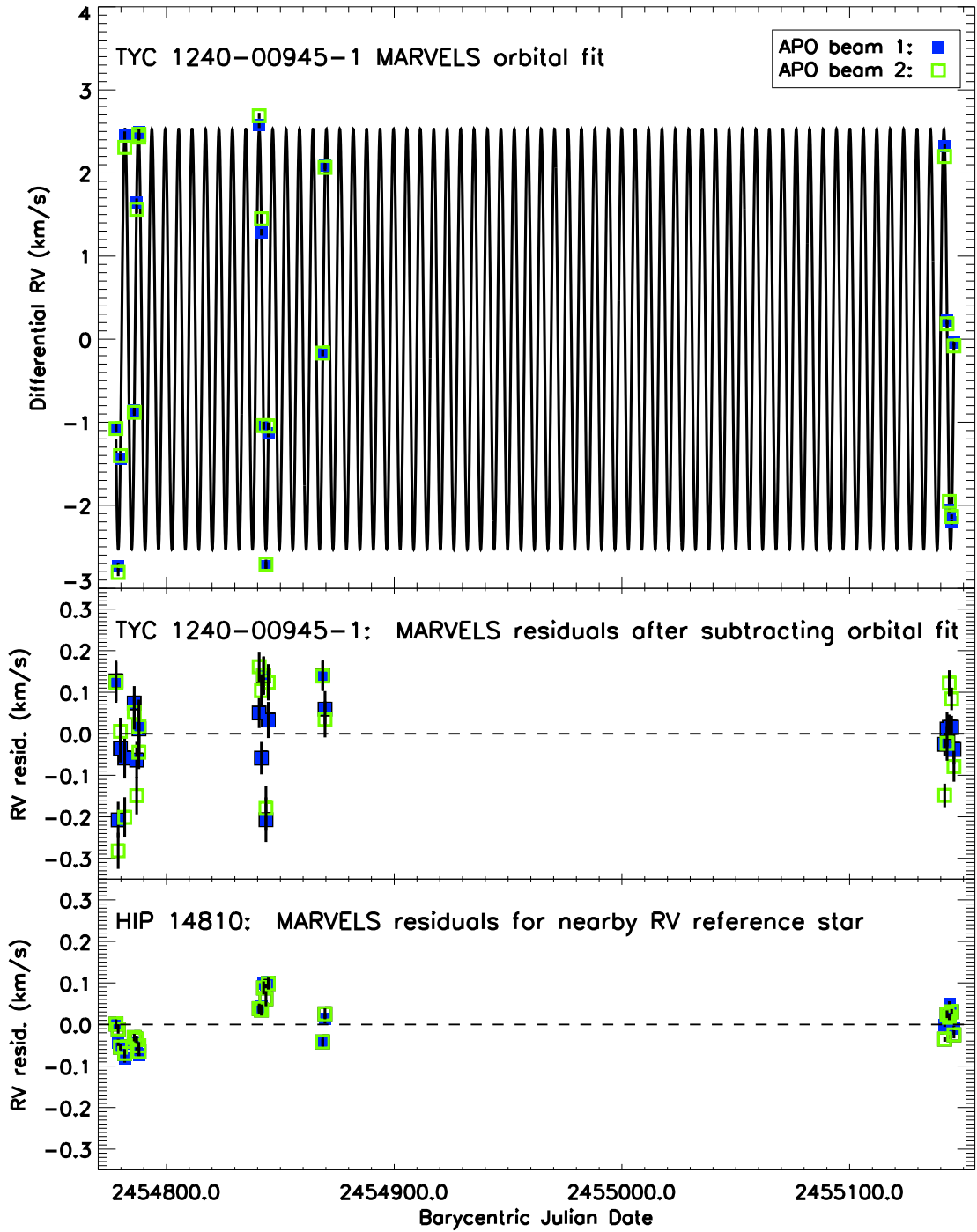}
\caption{\label{fig:rvcurve}\label{fig:rvresid} {\bf Top:}  MARVELS RV data and Keplerian orbital solution for TYC~1240-00945-1.  Beam 1 is shown with blue filled squares, and beam 2 with green open squares.  {\bf Center:}  The residuals for TYC~1240-00945-1, equal to the RVs from the top panel minus the orbital fit.  {\bf Bottom:}  The residuals for HIP~14810, a star with a known two-planet RV signal, observed through a nearby fiber during the same exposures as those plotted for TYC~1240-00945-1.}
\end{figure}

In Table \ref{tab:obsjournal}, we present the 20 radial velocities measured by the MARVELS instrument, and we show the RV curve as a function of time in Fig.~\ref{fig:rvcurve}.  Both beams are shown, and even though the error bars plotted in Fig.~\ref{fig:rvcurve} are photon-only and do not account for systematics (our procedure to determine more realistic error bars follows below), several of the beamwise pairs nonetheless agree within their error bars.

The MARVELS pipeline is still under development, and we find the RV scatter for other stars in the same field as TYC~1240-00945-1 (as well as for stars in other fields) is on average 2--3 times larger than the photon noise, on timescales greater than a month;  presumably, most stars observed are not astrophysically variable at this level, indicating that the excess scatter is due to systematics (note we will discuss the expected RV jitter for TYC~1240-00945-1 in Section \ref{sec:jitter}, since we need to determine the stellar properties first before searching for cases of similar stars in the literature).  
During pipeline development, we have examined the morphology of the RV residuals in the cases of several reference stars with known RV curves (either stable or planet-bearing) and found the systematic errors typically manifest in the form of month-to-month offsets at the level of tens of m~s$^{-1}$, such that the RV data within any individual month fits the known RV curve much better than over multiple months.  The offsets are often the same in direction and magnitude for both beams.
These systematic errors may be due to imperfections in the detailed preprocessing of the images, because we do not see these systematics at the same level when analyzing simulated stellar data free of real-world image distortions.  Since the exact factor by which the scatter exceeds the photon noise varies from star to star, we have decided to determine the excess scatter for the candidate at hand, to ensure that it falls in the typical range seen for other stars, and so is not responsible for the RV signal which we have interpreted as due to a companion. 

Our procedure for estimating the magnitude of the systematic errors
in the RV curve is as follows.  We assume that the systematic errors can
be well-modelled by applying a simple constant multiplicative scaling to the 
uncertainties derived from the photon noise alone.  We choose to use a multiplicative scaling of the error bars instead of adding a systematic error in quadrature to the statistical error bars because during pipeline development, we found the increase in RV scatter above the photon noise level is larger for fainter stars than brighter stars, so adding systematic error in quadrature would not be able to capture the overall form of the extra error as a function of signal-to-noise ratio.  
We designate this multiplicative scaling factor the ``quality factor'' $Q$.  We estimate $Q$ by performing a Keplerian fit to the RV dataset  (with the raw pipeline photon-noise uncertainties), allowing for a linear trend with time.  We then find the value of $Q$ such that the $\chi^2$/dof of the best-fit is equal to unity.

Following this error bar growth procedure, we found that the MARVELS
RVs for TYC~1240-00945-1 were affected by systematics at levels of
$Q_{\rm beam 1} = 2.21$ and $Q_{\rm beam 2} = 3.63$ for the two beams, respectively.  Multiplying the statistical error bars by $Q$, we get a median
scaled error bar of $92$\,m~s$^{-1}$ for beam 1 and $151$\,m~s$^{-1}$ for beam 2.  
After scaling the error bars, we performed a joint fit to beam 1 and beam 2 to provide a stronger constraint than a fit to either beam alone.  
The joint fit allows for different slopes and offsets between the two beams.  
This model is required because the two beams traveled through different parts of the instrument, and most importantly, experienced different optical path delays inside the interferometer (recall from Section \ref{sec:sdssobs} that there is a multiplicative factor that transforms fringe shift into radial velocity--  this factor depends on the delay).  
The parameters of this final joint MARVELS orbital
fit are given in Table \ref{tab:companionparams} below, and the fit is overplotted with the data in Fig.~\ref{fig:rvcurve}.  
The uncertainties were determined using the Markov Chain Monte Carlo (MCMC) method (see, e.g., \citealt{ford06}). 
Note the time is referenced to the time of inferior conjunction (i.e., the expected time of transit if the system is nearly edge-on), and is given as the Barycentric Julian Date (BJD) in the Barycentric Dynamical Time (TDB) standard \citep{eastman10}.  

The $Q$ values for the two beams are consistent with
that of a typical constant star's $Q$, $\sim \! 2-3$.  We
also checked the brighter planet-bearing reference star HIP~14810, which
was observed on the same plate at the same time.  Using the known
RV model \citep{Wright09}, we find this reference star has $Q_{\rm beam 1} = 5.29$ and $Q_{\rm beam 2} = 4.36$, with median statistical error bars
of $9.1$\,m~s$^{-1}$ for beam 1 and $10.4$\,m~s$^{-1}$ for beam 2.  The higher $Q$ for the brighter star is not an
especially surprising result, since systematic noise sources that are
independent of photon counts contribute a higher fraction of the total
error when photon noise is small.  Fig.~\ref{fig:rvresid} shows the
residuals of HIP~14810 relative to the model curve, on the same scale
as the RV residuals of TYC~1240-00945-1.  These residuals demonstrate
that we can recover the RV curve of a known planet-bearing star to a
level at least as good as our TYC~1240-00945-1 fit.  Hence, the level
of systematic uncertainty we find for TYC~1240-00945-1 is not unusual for its
field, and that level is small compared to the amplitude of RV variability
we find for TYC~1240-00945-1 and attribute to a companion--  MARVELS-1b.
\label{sec:errbarscaling}

\subsection{HET AND SMARTS RADIAL VELOCITY DATA}
 \label{sec:hetrv}

To further confirm that RV variability is indeed due to a companion, 
as well as to confirm the basic parameters of the orbital fit, we compared the RV observations obtained
from HET and SMARTS to those obtained by the MARVELS instrument.  
We found these RV data do verify the variability and periodicity, but the follow-up data sets comprise insufficient high quality data points to provide much additional refinement of the orbital fit parameters on top of the discovery data.

We first treated each RV dataset
independently, computing a separate orbital fit and estimating $Q$ for the dataset using the
procedure described above in \S\ref{sec:marvelsrv}.  This gives the minimal error bars that would be consistent with any
Keplerian orbital solution.  We use these separate fits only for
estimating the HET and SMARTS total error bars.

For the HET data we find $Q_{HET} = 15.3$, which is high, but expected
due to the preliminary nature of the pipeline used to reduce
the data (see \S\ref{sec:rvfollowup}).  We subtracted the RV model based on the MARVELS fit
from the HET points and found that the residuals could be fit by a
straight line (slope and offset) with $\chi^2 = 7.9$ and 7 degrees of
freedom.  Under the assumption that the errors are independent and normally
distributed, this corresponds to a 33.4\% probability of happening by
chance, so there is no evidence to reject the hypothesis that the HET
RVs are consistent with the MARVELS orbital fit.

For the SMARTS data we find $Q_{SMARTS} = 1.50$.  We subtracted the RV model based on the MARVELS fit from the SMARTS points and found that the residuals could
be fit by a straight line (slope and offset) with $\chi^2 = 20.4$ and
6 degrees of freedom.  Again assuming independent and normally distributed errors,
this has a 0.23\% probability of happening by chance, so there is
strong evidence to reject the hypothesis that the SMARTS RVs are
consistent with the MARVELS orbital fit.  However, given that the HET
and MARVELS RVs agree, we expect this discrepancy with the SMARTS
data merely reflects evidence for unidentified systematics in the SMARTS data, which is not surprising, given the
preliminary nature of the reduction of the SMARTS data (see
\S\ref{sec:rvfollowup}).

\begin{figure}[htbp] 
\begin{center}
\plotone{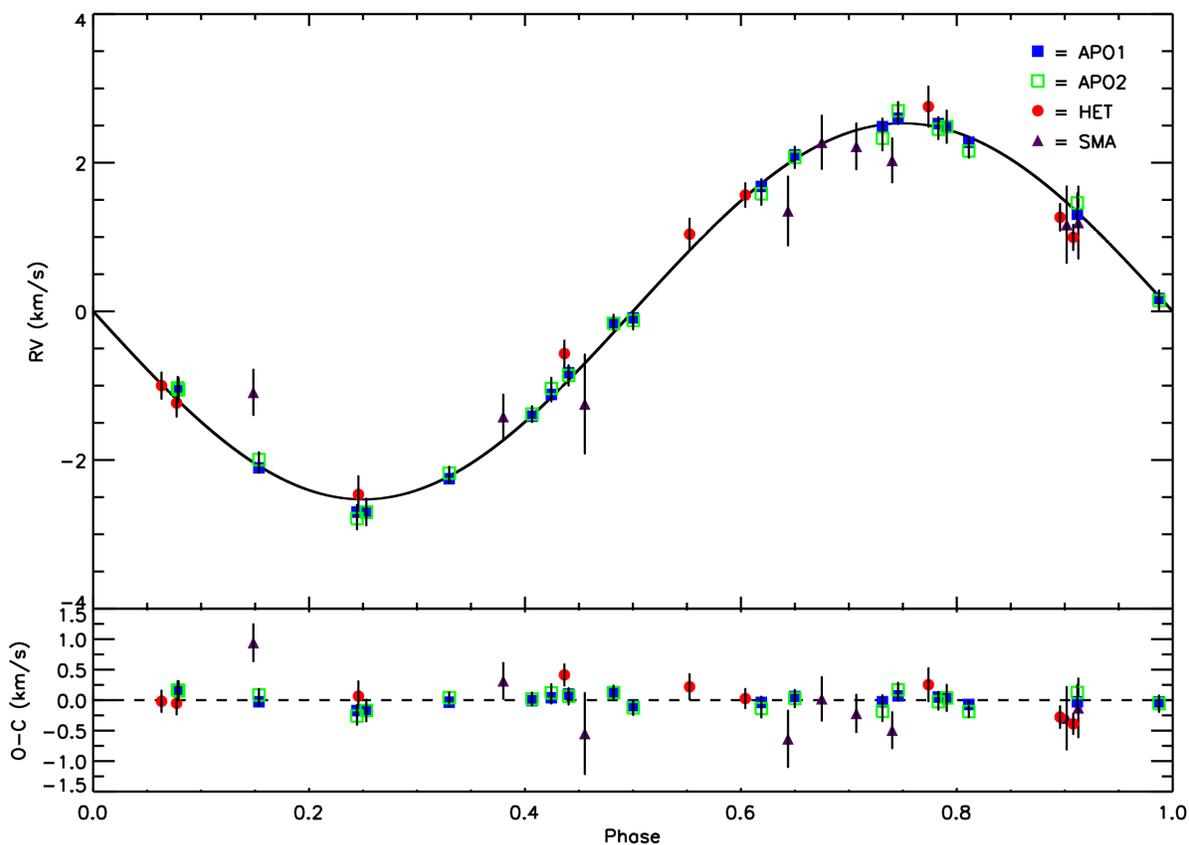}
\end{center}
\caption{Phase-folded Keplerian orbital solution and RV residuals for TYC~1240-00945-1.  Blue squares and green squares are MARVELS discovery data, red circles are HET data, and purple triangles are SMARTS data.  Error bars have been scaled up by the methodology in \S\ref{sec:errbarscaling}.  The bottom panel shows the residuals between the data points and the orbital solution. Note that
the HET and SMARTS data were not used in the Keplerian fit, and so provide an independent check
of the quality of the MARVELS data.}
 \label{fig:rvcurve_phased} 
\end{figure}

The four RV data sets are shown in Fig.~\ref{fig:rvcurve_phased}, phase-folded to the fitted period and phase (as determined from the fit to the MARVELS data alone).  This visually demonstrates the conclusion that the HET and SMARTS RV data confirm both the amplitude and phase of the variability.
We then tried an orbital fit to all three telescopes' data sets jointly, applying the same method that was used to jointly fit MARVELS beams 1 and 2, but now expanded to accommodate four RV data sets.  We found that the new period and amplitude derived, using all the data sets combined, matched the values adopted in Table \ref{tab:companionparams} to within the $1\sigma$ uncertainties, and furthermore, that the uncertainties themselves matched to within $\sim$10\%.

\section{MONITORING FOR PHOTOMETRIC VARIABILITY} 
 \label{sec:photmeas} 

\begin{figure}
\plotone{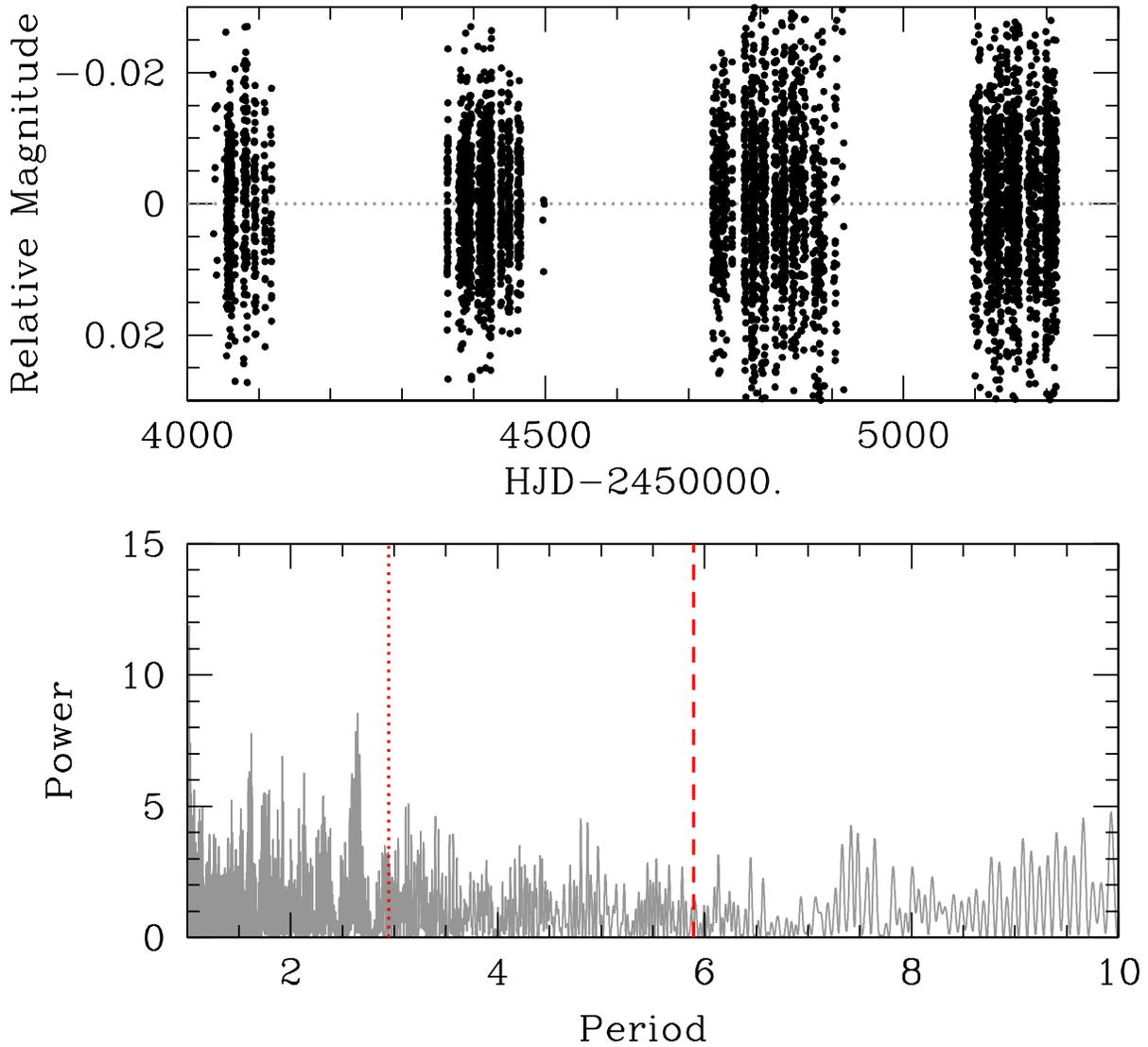}
\caption{\label{fig:kelt}
{\bf Top:}  KELT North light curve for TYC~1240-00945-1.  {\bf Bottom:}  Lomb-Scargle periodogram of the KELT data,
showing no evidence for any significant periodicities
for periods of $P=1-10$\,days, including the period of MARVELS-1b (vertical dashed line) and the first harmonic (vertical dotted line).}
\end{figure}

The KELT data for TYC~1240-00945-1 are displayed in Fig.~\ref{fig:kelt}, 
and show no evidence for variability.  The final weighted RMS is 0.92\%.  A weighted Lomb-Scargle
periodogram with floating mean \citep{lomb76,scargle82}
yields no significant peaks for periods of $1-10$\,d, and in
particular no evidence for any periodic variability near the period of
the companion or the first harmonic.  The improvement in $\chi^2$ for
a sinusoidal fit at the period of the companion is only $\sim 0.1$ relative to a constant flux fit.

\begin{figure}
\plotone{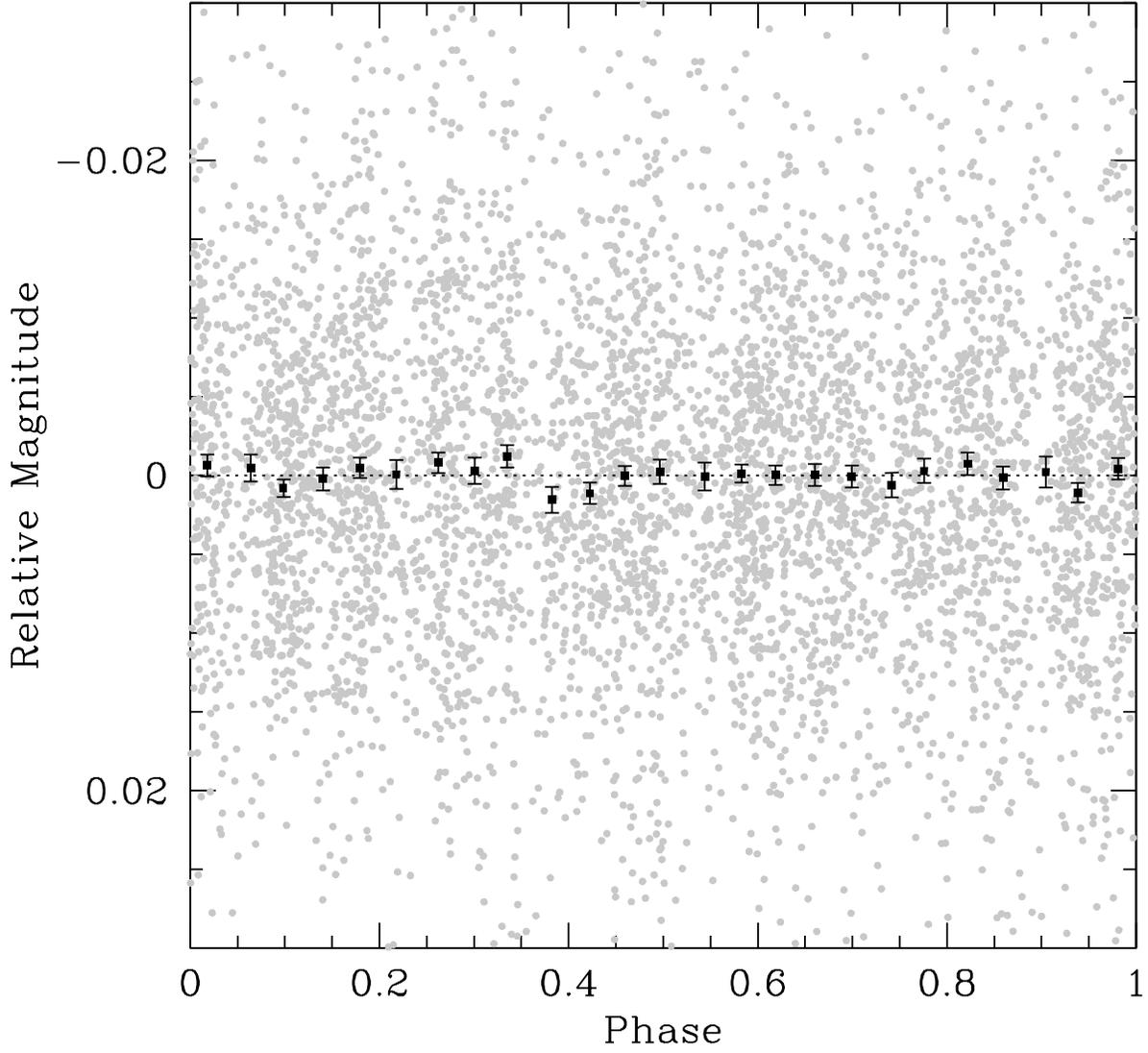}
\caption{\label{fig:binkelt}Grey points:  the KELT light curve for TYC~1240-00945-1, phased to the period of MARVELS-1b ($5.8953$\,days).  Black points:  the phased KELT light curve, binned using bin size $\Delta \phi = 0.04$.}
\end{figure}

Fig.~\ref{fig:binkelt} shows the KELT light curve phased to the
best-fit period of the companion ($5.8953$\,d), as well as the phased
light curve binned every 0.04 in phase (roughly the expected transit
duration for a mid-latitude transit).  The RMS of the binned curve is
0.059\%, with a $\chi^2$/dof of 0.85.  This is consistent with no
correlated (red) noise at the level of the RMS, since with an average of $\sim \! 200$ data points per phase bin, one would expect a factor of $\sim 15$ improvement for the binned RMS compared to the unbinned RMS.
We can also place an upper limit of 0.050\% on the maximum light curve
variability at a period half that of the period from the RV orbital
fit (at $\Delta \chi^2 = 9$), but this limit is insufficient to detect the
expected amount of ellipsoidal variability for this candidate system.
Using the equation in Table 2 of \citet{pfahl08}, we calculate
the ellipsoidal variation would only be 0.0019\% in amplitude.  Note the methods we use to calculate the physical parameters for the star and companion used in the equations in this section will be explained later, in \S\ref{sec:finaldet}.

We possess an ephemeris from the RV orbital fit to search for companion transits at the expected time.  However, prior to our exposition of the Monte Carlo analysis using the RV information, let us first consider approximately what S/N to expect, calculated under the simplifying assumption of a random ephemeris (allowing us to write an {\it analytic} expression for the S/N).  Based on the semimajor axis of $a = 0.071$\,AU for an edge-on system, the {\it a priori} transit probability for the companion is fairly high, $R_*/a=14.4$\%.  The expected duration of a central transit is $\sim\!R_*P/(\pi a)=6.49$\,hours, and the expected depth is $\delta\!\sim\!(r/R_*)^2=0.218\%(r/R_{Jup})^2$, where $r$ is the radius of the companion.  Using these values, the expected S/N of a transit in the KELT data can be estimated,
\begin{equation}
{S/N} \sim N^{1/2} \left(\frac{R_{*}}{\pi a}\right)^{1/2} \frac{\delta}{\sigma}\sim 3.5 \left(\frac{r}{R_{Jup}}\right)^2
\label{eqn:transitsnr}
\end{equation}
where $N=5036$ is the number of data points, and $\sigma \! \sim \! 1\%$ is the typical uncertainty.  Thus the detection of a transit using KELT data is challenging if the radius of the companion $r \! \lesssim \! R_{Jup}$, as is expected based on the likely age of the star (\S\ref{sec:tidal}) and the minimum mass of the companion \citep{baraffe03}.

Detailed limits on transits are produced by using the same Monte Carlo analysis as described in \citet{fleming10} to incorporate our transit ephemeris from the RV data.  Briefly, we use the distribution of companion periods and expected transit times from the MCMC chain derived from the fit to the MARVELS RV data (\S\ref{sec:marvelsrv}) to predict a distribution of transit times in the KELT data. For each link of the MCMC chain, we consider the uncertainty in the inferred radius of the primary due to the uncertainties in the spectroscopically measured $T_{\rm eff}$, $\log{g}$, and [Fe/H] (see \S\ref{sec:finaldet}), and we also consider a uniform range of transit impact parameters.  For a given assumed radius for the companion, for each link we can then compute the expected transit curve using the routines of \citet{mandel02}, which are fit to the KELT dataset, computing the difference in $\chi^2$ relative to a constant flux fit to the data.  This is repeated for each link in the Markov chain, as well as for a variety of different companion radii.  We find that our best-fit transit light curve has $\Delta\chi^2 \! \simeq \! -5$ relative to a constant flux fit.   Based on analysis of the noise properties of the KELT light curve and the number of trials we performed searching for a transit, we estimate that $\Delta \chi^2 \! \la \! -16$ is generally indicative of a reliable detection, and thus this improvement is not significant.

\begin{figure}
\plotone{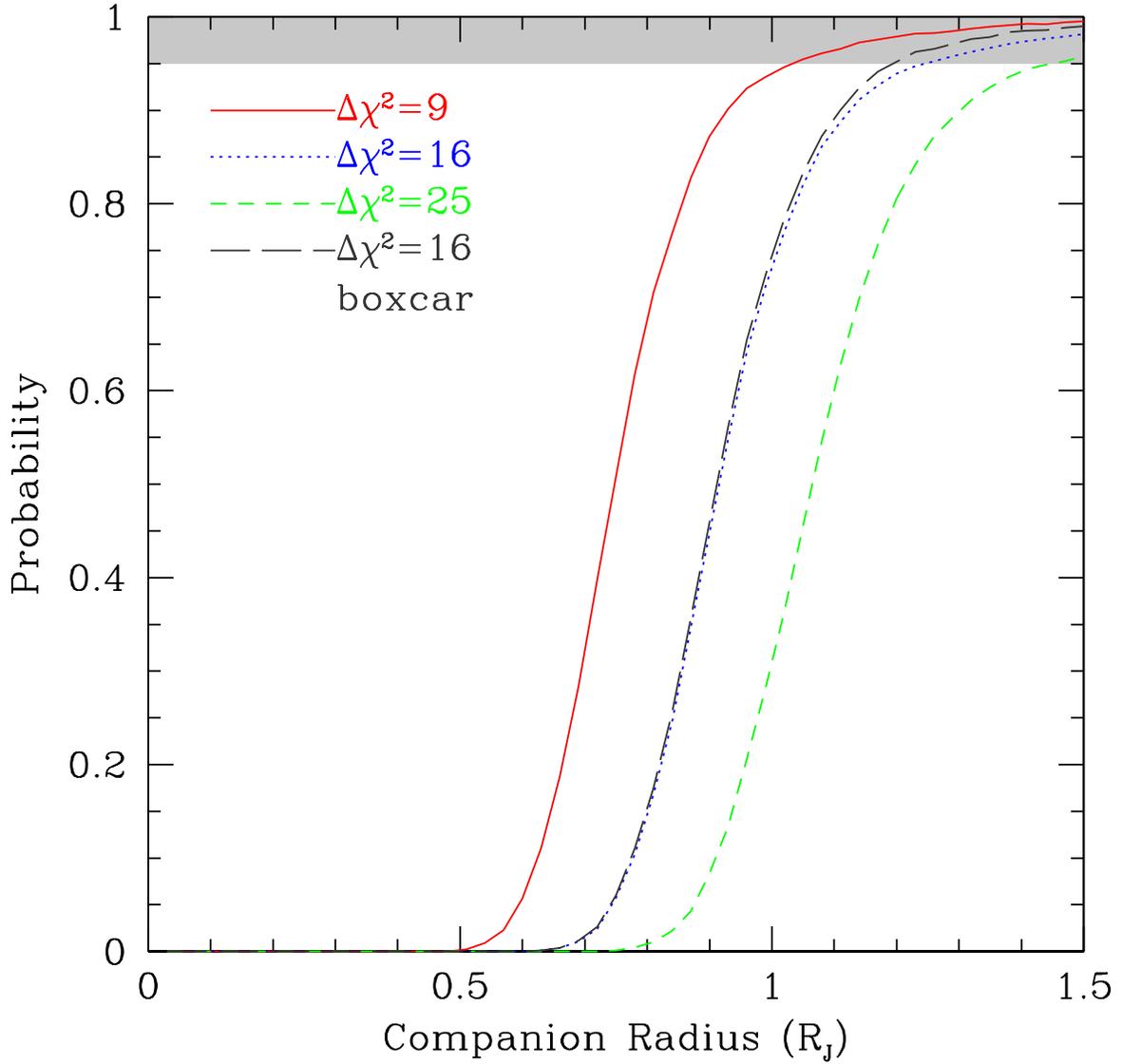}
\caption{\label{fig:exctrans}
Probability that transits of MARVELS-1b are excluded
at levels of $\Delta\chi^2=9$ (solid red), 16 (dotted blue), and 25 (dashed
green),  based on the analysis of the KELT photometric dataset, as a function of the radius
of MARVELS-1b.  Also shown is the case for $\Delta\chi^2=16$, but assuming a
box-shaped transit (black, long dashes) instead of a limb-darkened light curve model.}
\end{figure}

We then determine the fraction of trials that lead
to a $\Delta \chi^2$ greater than some threshold value.  The results
for $\Delta\chi^2$=9, 16, and 25 are shown in Fig.
\ref{fig:exctrans}.  We find that $\sim \! 95$\% of MCMC realizations of transit models for companion radii $> \! 1.2\,R_{Jup}$ lead to fits to our light curve that are
excluded by our data, in the sense of producing a $\Delta\chi^2$ that
is worse by more than 16 relative to a constant fit. Therefore, we
exclude with $\sim\!95$\% confidence that the companion transits if
it has a radius larger than $\sim \! 1.2\,R_{Jup}$, and with $\sim \! 75$\%
confidence if it has a radius larger $\sim \! R_{Jup}$.  We conclude
that while transits of a Jupiter-radius companion are
unlikely, they are not definitively excluded.

\section{STELLAR PARAMETERS} 
 \label{sec:params} 

We have made multiple determinations of the stellar parameters of the
host star, using several different sets of data and analysis methods, described below.
The results are summarized in Table \ref{tab:stellarparamsraw}.
We note that, although the different determinations are generally mutually consistent, the
uncertainties associated with each are simply formal statistical uncertainties, which 
have not been externally
calibrated.  We expect that these formal uncertainties are likely 
underestimates of the true uncertainties.  Therefore, we conservatively choose to report the median of the
three highest resolution spectroscopic results as our best estimate of
the stellar parameters, and take the uncertainty as the standard
deviation of the three estimates.
The final stellar
parameters we adopt are effective temperature $T_{\rm eff} = 6186 \pm 92$\,K, surface gravity $\log g = 3.89 \pm 0.07$\,(cgs), and metallicity $\rm [Fe/H] = -0.15 \pm 0.04$.  These
and other properties of the star are listed in Table
\ref{tab:stellarparams}.

\subsection{Fitting of Spectral Lines}

We analyzed the
extracted APO 3.5-m spectra to determine the stellar properties
in a careful hand-guided analysis according to
the techniques used by \citet{Laws2003}, which are described more
fully (excepting recent improvements) in \citet{Gonzalez1998}.
Briefly, we make use of the line analysis code MOOG
(\citealt{sneden73}, updated version), the \citet{kurucz93} LTE
plane-parallel model atmospheres, and equivalent width (EW)
measurements of 62 Fe I and 10 Fe II lines to determine the
atmospheric parameters $T_{\rm eff}$, $\log g$, microturbulence $\xi_t$, and [Fe/H].
The formal uncertainties were calculated using the method in \cite{Gonzalez1998}.  The values are listed in Table
\ref{tab:stellarparamsraw}.

As a check, we performed a second analysis of the APO spectra using
the code Spectroscopy Made Easy \citep[SME; see][]{Valenti05}. SME is
an IDL-based program that uses synthetic spectra and least-squares
minimization to determine the stellar parameters (e.g., $T_{\rm eff}$,
$\log\,g$, [Fe/H], $v\,\sin\,i$, etc.) that best fit an
observed spectrum. To constrain the stellar parameters, we analyzed
three wavelength regions ($5160-5190$\,\AA, $6000-6200$\,\AA, and
$6540-6590$\,\AA) used by \citet{Stempels2007}. The first
region is sensitive to $\log\,g$. The second region contains a large
number of spectral features of different elements, and is sensitive to
[M/H] and $v\,\sin\,i$. The third region contains
H$\alpha$, and the broadening of the outer wings of this line is
sensitive to $T_{\rm eff}$. We fitted all three regions simultaneously
using SME to estimate the stellar parameters of TYC~1240-00945-1.  SME
was unable to determine $v\,\sin\,i$ to a level finer than the velocity
resolution of the APO 3.5-m spectra ($\sim \! 9$\,km~s$^{-1}$ at $R \! \sim
 \! 31500$).  We derived parameters that agreed with those determined from the same spectra using the \citet{Laws2003} methodology.  The values are listed in Table
\ref{tab:stellarparamsraw}.

The stellar parameters were also verified using the ESO 2.2-m FEROS spectra.  Measurements of the equivalent widths were
carried out automatically using the ARES code \citep{Sousa07}.
Given the
high S/N and broad spectral range of the spectrum, results 
were obtained for a large number of atomic lines. 
However, after a careful inspection, only 21 Fe I and 9 Fe II lines (from the list in Table 2 of \citet{Ghezzi10}) were considered sufficiently reliable to be used in the determination of the stellar parameters. Applying the technique described in \citet{Ghezzi10}, the following results were obtained: $T_{\rm eff}=6186 \pm 82$\,K, $\log g = 4.01 \pm 0.17$, $\xi_t = 1.26 \pm 0.17$\,km~s$^{-1}$, and [Fe/H]$ = -0.14 \pm 0.08$, where the
formal uncertainties were calculated as in \cite{Gonzalez1998}.

The projected rotational velocity of TYC~1240-00945-1 was estimated from the high-resolution FEROS spectrum using a technique similar to the one described in \citet{Ghezzi09}. The expectation from FEROS simulations is that the high oversampling of the line spread function for the FEROS spectrum allows us to probe to much lower $v\,\sin\,i$ than achievable with the APO 3.5-m
spectra, even though the FEROS resolving power is only moderately higher.  We measure $v\,\sin\,i$ by simultaneously fitting the
macro-turbulence velocity and $v\,\sin\,i$ for three moderately
strong Fe I spectral lines. A grid of synthetic spectra was
generated, varying $v\,\sin\,i$, the macro-turbulence velocities and
the adopted [Fe/H], the latter by 0.05 dex around the mean value
given above. Small adjustments in the continuum level under 0.4\%
were allowed, to account for possible errors in the normalization
process. In addition, small shifts in the central wavelengths of
the Fe I lines were needed in order to properly match the observed
lines. Values for $v\,\sin\,i$ and macro-turbulence were determined
separately for each of the Fe I lines considered, based on standard
reduced-$\chi^2$ minimization. The results obtained for the three Fe I
lines were consistent, yielding a $v\,\sin\,i$ in the range $1.1-3.2$\,km~s$^{-1}$, and macro-turbulence in the range $4.5-4.7$\,km~s$^{-1}$. 
The latter values are in good agreement with the macroturbulence velocity derived from Equation 1 in \cite{Valenti05} and $T_{\rm eff}=6186$\,K.
Our best
estimate for $v\,\sin\,i$ was computed as the mean of the three
values, yielding $v\,\sin\,i=2.2 \pm 1.5$\,km~s$^{-1}$, where the uncertainty is
the RMS value;  this RMS scatter is approximately equal to the intrinsic uncertainty
of the fitting procedure, which is typically $1-2$\,km~s$^{-1}$.  Note that when we tried recovering $v\,\sin\,i$ from simulations of FEROS spectra at the S/N of the TYC~1240-00945-1 spectrum, we found that even lower $v\,\sin\,i$ would indeed be detectable at the FEROS resolution. 
However, as discussed below in \S\ref{sec:tides}, the lower 1$\sigma$ limit
of 0.7\,km~s$^{-1}$ leads to a very long rotation period which is astrophysically unlikely;  it is more probable that the true $v\,\sin\,i$ lies within the upper half of the estimated range from the fit.

\begin{figure}
\plotone{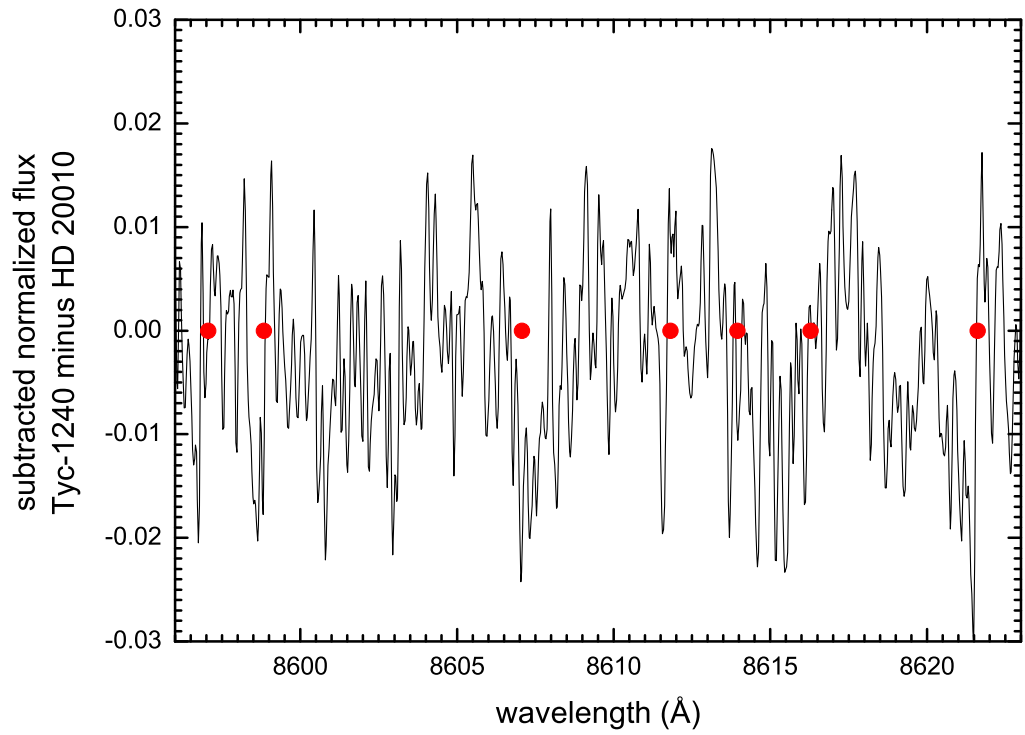}
\caption{\label{fig:specdiff}
The normalized spectrum of TYC 1240-00945-1 minus the normalized spectrum of HD 20010 (an F8IV star with similar stellar parameters), in the wavelength range 8592-8626\AA.  Locations of some spectral lines of a late F subgiant are indicated by red dots.}
\end{figure}

We searched the FEROS spectra for any indication of spectral features from a secondary star blended with the primary, as might be expected if the RV signal were in fact caused by a nearly pole-on orbit of a low-mass stellar companion.  To make a quantitative search for extra flux, we computed the difference of the normalized spectrum of TYC 1240-00945-1 with a template FEROS spectrum of the primary star of the binary HD 20010, a well-studied F subgiant \citep[][]{balachandran90, santos04, luck05} with stellar parameters similar to those we derived for TYC 1240-00945-1.  We examined the $8570-8630$\,\AA~ region, where the spectrum has good continuum level determination, several spectral features (mostly due to Fe I), and where the contrast ratio between an M dwarf and the primary would be relatively high, before the red-end fall-off in detection efficiency of the FEROS spectra (in this range, the S/N per pixel of TYC~1240-00945-1 and HD 20010 were high:  180 and 390, respectively).  From the \citet{pickles98} low resolution spectral library, we computed the expected ratio of fluxes between F8IV and M0V stars over this wavelength range to be 1.5\%.  We would expect that ratio to manifest as a difference in line ratios between the template and target spectra, with the M dwarf's flux filling up the cores of the F star's lines.  However, the difference spectrum shows no detectable systematic offsets at the locations of HD 20010's lines;  rather, the difference is evenly distributed around zero, with a standard deviation of 1.0\%.  The difference spectrum is shown in Fig.~\ref{fig:specdiff}. This amount of deviation is expected since there is uncertainty in picking a template which would exactly match TYC~1240-00945-1.  Thus there is no evidence for an M0V contaminating spectrum, although much cooler M dwarfs would provide less than 1.5\% contaminating flux and might not be visible given the noise in our measurement.

\subsection{Spectral Energy Distribution Fitting}
\label{sec:sedfit}

As an additional check on the parameters of TYC~1240-00945-1, 
we performed a model atmosphere fit to the observed spectral energy distribution
(SED) from the optical fluxes from HAO (\S\ref{sec:obsphotometry}) and near-IR fluxes from 
2MASS \citep{skrutskie06}. The absolute photometric measurements in the $g'r'i'JHK_S$ passbands 
(see Table \ref{tab:stellarparams}) were converted to physical fluxes using the published 
SDSS\footnote{\url{http://www.sdss.org/dr7/algorithms/fluxcal.html}}
and 2MASS\footnote{\url{http://ssc.spitzer.caltech.edu/documents/cookbook/html/cookbook-node207.html}} 
zero-points, together with published color-dependent corrections to the 
passband effective wavelengths \citep{moro00}. 
The model atmospheres used in the fitting are the NextGen atmospheres of 
\citet{hauschildt99},
which are gridded in $T_{\rm eff}$ by 100\,K, in $\log g$ by 0.5\,dex, and in
[Fe/H] by 0.5\,dex. 
We performed a least-squares fit of this model grid to the six flux measurements,
with the extinction $A_V$ and the overall flux normalization as additional free 
parameters.

\begin{figure}[htbp] 
\begin{center}
\plotone{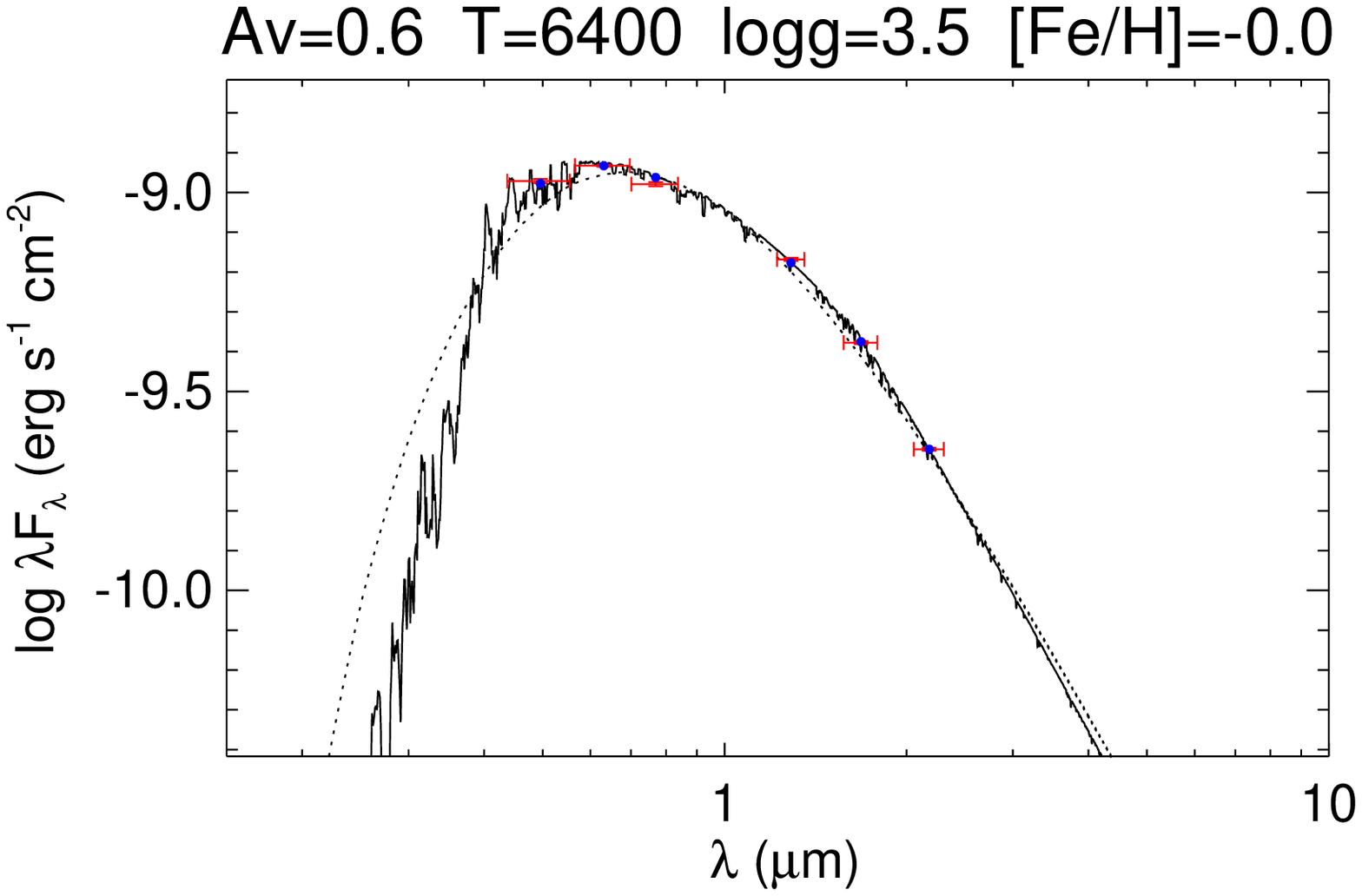}
\plotone{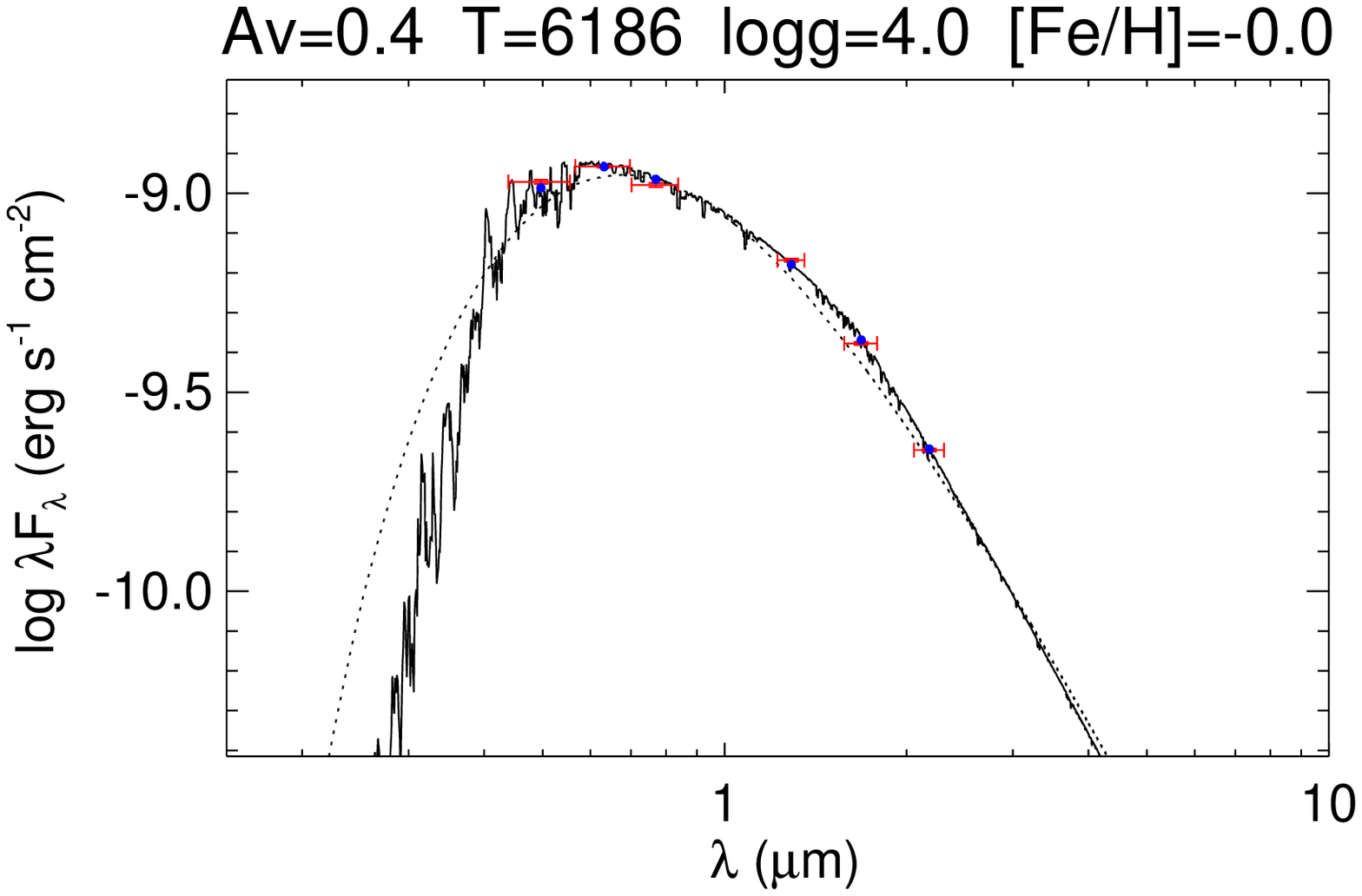}
\end{center}
\caption{\label{fig:sedallvarying} {\bf Top:}  NextGen model atmospheres SED fit to the color photometry of TYC~1240-00945-1, allowing all parameters to vary.  \label{fig:sedspecfixed}  {\bf Bottom:}  SED fit to the color photometry of TYC~1240-00945-1, with only $A_V$ and the normalization as variables ($T_{\rm eff}$, [Fe/H], and $\log g$ locked).}
\end{figure}

We initially allowed all of the variables-- $T_{\rm eff}$, $\log g$, [Fe/H],
$A_V$, and flux normalization-- to be fit as free parameters. 
We limited the $A_V$ to a maximum of 0.65, corresponding to the maximum 
line-of-sight extinction as determined from the dust maps of \citet{schlegel98}.
The resulting fit is shown in Fig.~\ref{fig:sedallvarying}, with $T_{\rm eff}=6400^{+400}_{-600}$\,K,
$A_V=0.6^{+0.05}_{-0.45}$, $\log g = 3.5 \pm 1.5$, and [Fe/H] $= 0.0 \pm 2.0$.

These values are consistent with those derived spectroscopically.  
However, the available photometry does not strongly constrain the stellar
parameters as there is a
very strong degeneracy in the SED fit between $T_{\rm eff}$ and $A_V$, due to
the lack of absolute flux measurements at wavelengths bluer than $0.5$\,$\mu$m.
Thus, we re-fit the fluxes with $T_{\rm eff}$ fixed at the spectroscopic value
of $6186$\,K, [Fe/H] fixed at $0.0$, and $\log g$ fixed at $4.0$;  the only 
remaining free parameters are $A_V$ and the normalization.  In this way we
use the photometry to strongly constrain the line-of-sight extinction.
The resulting best fit, with $A_V=0.40 \pm 0.05$, is displayed in Fig.~\ref{fig:sedspecfixed}. 

Adopting this $A_V$, which implies $E(B-V)=0.13$ using the reddening law of 
\citet{bessell88}, we can check $T_{\rm eff}$ from the broadband colors alone, using the recent color calibrations of \citet{casagrande10}.
For example, from the $J-K_S$ color we find $T_{\rm eff}=6147$\,K, while from
the $V-K_S$ color we obtain $T_{\rm eff}=6299$\,K.  Thus, given a reasonable estimate of $A_V$, even when we use individual colors instead of fitting them all simultaneously, the $T_{\rm eff}$ estimates are consistent with the spectroscopically determined value to within $\sim \! 100$\,K.

\subsection{Final Determination of the Stellar Parameters and Companion Parameters}
\label{sec:finaldet}

We determine the mass and radius of the parent star, TYC~1240-00945-1, from $T_{\rm
eff}$, $\log g$, and [Fe/H] using the empirical polynomial relations
of \citet{Torres2010}, which were derived from a sample of eclipsing
binaries with precisely measured masses and radii.  We estimate the
uncertainties in $M_*$ and $R_*$ by propagating the uncertainties in $T_{\rm
eff}$, $\log g$, and [Fe/H] (see Table \ref{tab:stellarparams}) using
the covariance matrices of the \citet{Torres2010} relations kindly provided by
G. Torres.  Also, since the polynomial relations of \citet{Torres2010} were derived empirically, the relations were subject to some intrinsic scatter, which we add in quadrature to the uncertainties propagated from the stellar parameter measurements.  The final stellar mass and radius values we obtain in this way
are $M_* = 1.37 \pm 0.11\,M_{\odot}$ and $R_* = 2.20^{+0.25}_{-0.22}\,R_{\odot}$.

Using the derived value of $M_*$, we estimate a minimum
mass (i.e., for $\sin\,i=1$ where $i$ is the orbital inclination) for the companion, MARVELS-1b, of $m_{\rm min}=28.0 \pm 1.5\,M_{Jup}$,
where the uncertainty is dominated by the uncertainty in the primary mass.  
In fact, the mass function,
\begin{equation}
 \frac{(m\sin i)^3}{(M_*+m)^2}\,\propto\,K(1-e^2)^{1/2}P^{1/3},
\label{massfunc}
\end{equation}
is more precisely determined.  We find $(m\sin i)^3/(M_*+m)^2 = (9.75 \pm 0.32) \times 10^{-6}\,M_\odot$.
With our adopted value of $M_*$, we can also estimate the semimajor axis
$a=0.071 \pm 0.002\,{\rm AU}$, assuming an edge-on orbit;  for less inclined orbits, the semimajor axis is larger.  

The small minimum mass of the companion positions it as a good
short-period brown dwarf desert candidate.  In order for it to be a
low-mass star rather than a brown dwarf, the orbital inclination would
have to be close to face-on.  In order to explore further the probability
that the companion has a mass greater than the hydrogen burning limit,
we conducted a Bayesian analysis to estimate the posterior probability
distribution for the companion mass, using the methodology described in
Section 7 of \citet{fleming10}:  an MCMC chain is constructed starting from a distribution of stellar parameters and error bars as adopted for TYC~1240-00945-1 in Table \ref{tab:stellarparams}, stellar masses are determined using \citet{Torres2010}, and companion masses are determined using a random distribution of inclinations.
This analysis assumes
a uniform distribution in $\cos\,i$, includes uncertainties on
the orbital and host star parameters, and adopts priors on the luminosity
ratio and mass ratio for the companion.

Of course, the posterior distribution of the true companion mass
depends on our adopted prior for the companion mass ratio
distribution (e.g., \citealt{ht10}).  Given that few brown dwarf
companions are known, the constraints on the companion mass ratio distribution in the
mass regime of interest are poor.  Indeed, this is what
makes this object interesting, and this distribution is precisely what
we would like to infer from a larger ensemble of similar detections.
Nevertheless, we can adopt various simple and plausible forms for the
mass ratio distribution, and then use these to infer posterior
probability distributions for the true mass.  From Doppler surveys for
exoplanets, it is known that Jupiter-mass companions are significantly
more common than brown dwarf companions, and that the frequency of
planetary companions declines for larger masses, such that the mass
function is roughly uniform in the logarithm of the planet mass for
$m \la 10~M_J$ \citep{cumming08}.  It is not known if this form holds
for companions with mass significantly larger than $\sim 10~M_J$, but
it is clear that the frequency of companions in the brown dwarf regime must
reach a minimum at some point and then rise again, given that M dwarf
companions with masses just above the hydrogen burning limit are known
to be more common than brown dwarf companions. \citet{Grether06} found that
this minimum (the driest part of the brown dwarf desert) occurs at a
companion mass of $31_{-18}^{+25}~M_J$. Thus the minimum mass of MARVELS-1b is near the minimum of the companion
mass function, and prior mass ratio distributions that are falling,
flat, or perhaps rising shallowly in $\log{q}$ are all
equally plausible (see Figure 11 of \citealt{Grether06}).

\begin{figure}[htbp] 
\begin{center}
\plotone{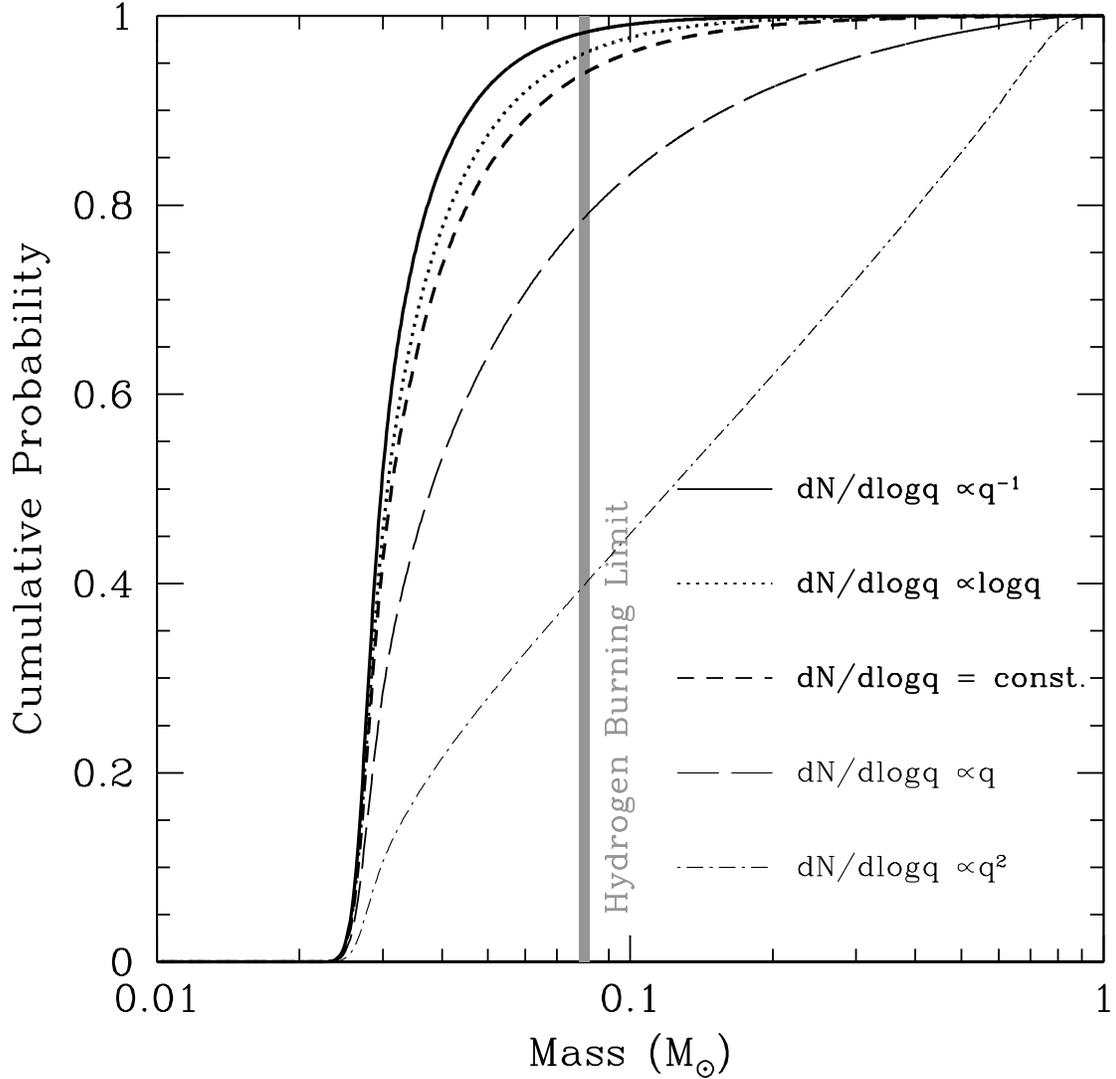}
\end{center}
\caption{\label{fig:bayes} Cumulative probability that the mass of MARVELS-1b is less than a given mass, in units of solar masses.  These probabilities account for the uncertainties and covariances between the parameters of the Keplerian orbital fit, the uncertainty in the host star mass, the assumption of a uniform distribution of $\cos\,i$, and the adoption of five different priors for the distribution of companion mass ratios ${\rm d}N/{\rm d}\log{q}$.}
\end{figure}

We therefore consider five different priors on the companion mass ratio
distribution:  ${\rm d}N/{\rm d}\log{q} \propto q^{-1}$, $\propto
\log{q}$, constant, $\propto q$, and $\propto q^2$.  The first three
are falling or constant with $\log{q}$, and the latter two are rising with
$\log{q}$.  From the results of \citet{Grether06}, we believe the first
three are the most plausible, while the first four almost certainly
bracket the likely range of distributions for companions close to the
relevant regime.  The resulting cumulative probabilities for the
companion mass for the five different priors are plotted in Figure
\ref{fig:bayes}. For the three favored priors, we conclude that at
$\ga 90\%$ confidence the actual mass is below the hydrogen-burning
limit.  For the prior that is uniform in (linear) mass ratio, ${\rm
d}N/{\rm d}\log{q} \propto q$, there is a $\sim 25\%$ probability that
the companion is in fact a low-mass star, whereas it is only for the
assumption of relatively steeply rising mass ratio distribution
(${\rm d}N/{\rm d}\log{q} \propto q^2$) that
the companion is more likely to be a star.  Again, we do not believe
such a distribution is very likely to be correct for this regime of companion
mass, but given the poor constraints, we cannot absolutely
exclude it either.  Finally, we note that for the last two priors, the
precise form of the posterior distribution depends on our
imposed constraint on the luminosity ratio, which is somewhat
uncertain.

With a reddening of $E(B-V)=0.13$ (\S\ref{sec:sedfit}), the system is evidently seen much of the way
through the full reddening along this line of sight, which from the \citet{schlegel98} dust maps is $E(B-V)=0.186$. 
The physical distance of the system can be estimated from its luminosity
and apparent magnitude. First we compute the bolometric magnitude of the star as
$M_{\rm bol} = 4.74 - 2.5 \log (L/L_\odot)$, where 4.74 is the bolometric
magnitude of the Sun. 
The luminosity is calculated from the Stefan-Boltzmann law applied to the
$T_{\rm eff}$ and stellar radius calculated above, and we adopt a $BC_V = -0.17$ as appropriate for
its spectral type (e.g., \citealt{kenyon95}).
The absolute magnitude is therefore 2.91.
Adopting $A_V=0.4 \pm 0.05$ (\S\ref{sec:sedfit}), this yields a distance $d=280 \pm 30$\,pc.

\subsection{Expected Stellar RV Jitter}
\label{sec:jitter}

Starspots and motions of the stellar surface are possible astrophysical sources of noise that can interfere with searches for companion RV signals.  These sources are commonly referred to as ``jitter'', and are explored by, e.g., \citet{saar98}, \citet{wright05}, \citet{lagrange09}, and \citet{isaacson10}.  For late F dwarfs of $B-V > 0.5$, they find typical jitters in the $\sim$10\,m~s$^{-1}$ range, with the most extreme outliers at $\sim$100\,m~s$^{-1}$.

TYC~1240-00945-1 is slightly evolved, so one wonders whether it might experience larger jitter than for F dwarfs.  However, it still lies at $B-V$ and $M_V$ below and redward of the instability strip \citep[for a review of the position of the strip, see, e.g.,][]{sandage06}, and shows no signs of activity based on the time-series photometry (Section \ref{sec:photmeas})--  so one shouldn't expect multi-periodic pulsations at the level of, e.g., the $\sim$400\,m~s$^{-1}$ RV jitter of the brown dwarf-hosting, instability strip, A9V star HD 180777 \citep{galland06}.  
Rather, F stars with stellar parameters similar to TYC~1240-00945-1 can be fairly quiet in terms of the RMS scatter attributable to RV jitter:  $\sim$4-5\,m~s$^{-1}$ in the case of the F6 star HD 60532 \citep{desort08}, and $\sim$10\,m~s$^{-1}$ in the case of the F7 star HD 89744 \citep{korzennik00}.

We conclude that the levels of RV jitter expected for this combination of stellar parameters are too low to be responsible for the $K=2.533\pm0.025$\,km~s$^{-1}$ of the TYC~1240-00945-1 RV signal, although they could be a contributor to the extra error we have regarded as systematics in the RV analysis.

\section{DISCUSSION}

\subsection{Evolutionary state of the host star}
\label{sec:tidal}

\begin{figure}
\plotone{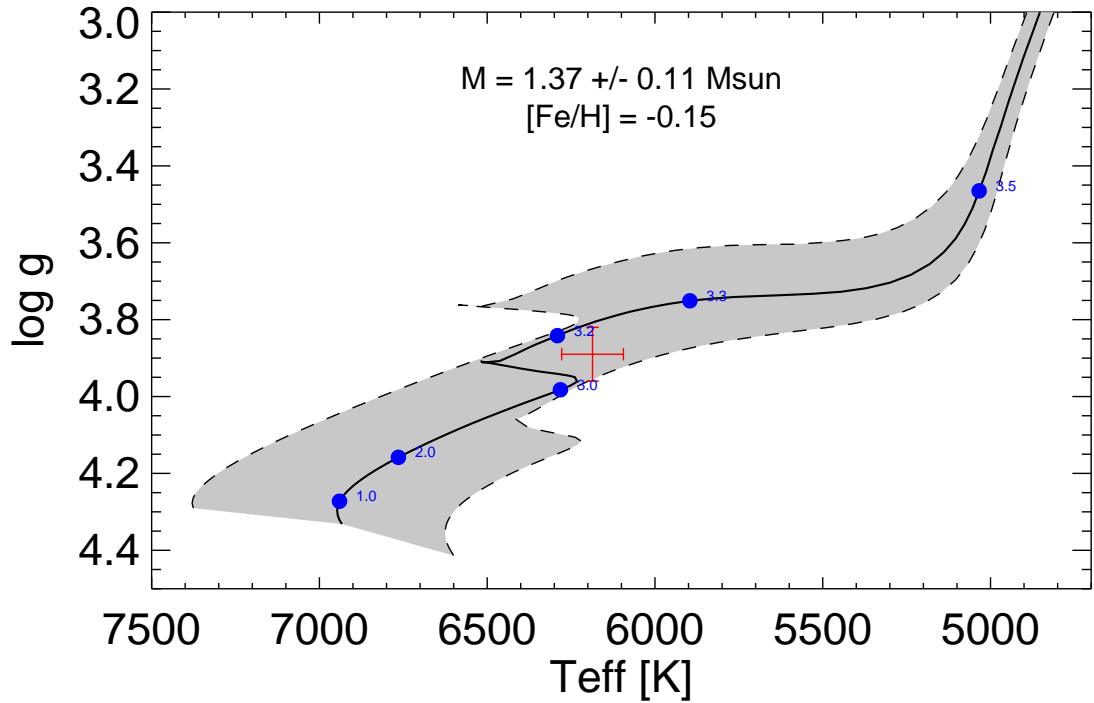}
\caption{\label{fig:evolution}The evolutionary track for an object with $M=1.37 \pm 0.11\,M_\odot$, at [Fe/H]=-0.15.  Ages of 1.0, 2.0, 3.0, 3.2, 3.3, and 3.5\,Gyr are indicated as dots.  The possible tracks for up to a $1\sigma$ deviation in the mass are shown by the shaded region.  The stellar parameters for TYC~1240-00945-1, with $1\sigma$ error bars, are shown by the cross.}
\end{figure}

In Fig.~\ref{fig:evolution} we compare the spectroscopically measured $T_{\rm eff}$ 
and $\log g$ of TYC~1240-00945-1 (red error bars) against a theoretical stellar evolutionary 
track from the Yonsei-Yale (``Y$^2$") model grid
(see \citealt{demarque04} and references therein).  
The solid curve represents the
evolution of a single star of mass $1.37\,M_\odot$ (the mass of TYC~1240-00945-1
inferred from the empirical calibration of \citealt{Torres2010}; see above) 
and metallicity of [Fe/H]=$-0.15$ (as determined spectroscopically), 
starting from the zero-age main sequence (lower left corner), across the 
Hertzsprung gap, and to the base of the red-giant branch. 
Symbols indicate various time points along the track, with ages in Gyr labeled.
The dashed curves represent the same evolutionary track
but for masses $\pm 0.11\,M_\odot$, representative of the $1\sigma$
uncertainty in the mass from the \citet{Torres2010} relation. The filled gray region
between the mass tracks therefore represents the expected location of a star
of TYC~1240-00945-1's mass and metallicity as it evolves off the main sequence. We
emphasize that we have not directly measured the mass of TYC~1240-00945-1, and thus
we are not attempting to test the accuracy of the stellar evolutionary tracks.
Rather, our goal is to use these tracks to constrain the evolutionary status of the TYC~1240-00945-1 system. 

The spectroscopically measured $T_{\rm eff}$, $\log g$, and [Fe/H] place
TYC~1240-00945-1 near the beginning of the subgiant phase, just prior to crossing
the Hertzsprung gap to the base of the red giant branch, with an estimated
age of $\sim \! 3$\,Gyr.

\begin{figure}
\plotone{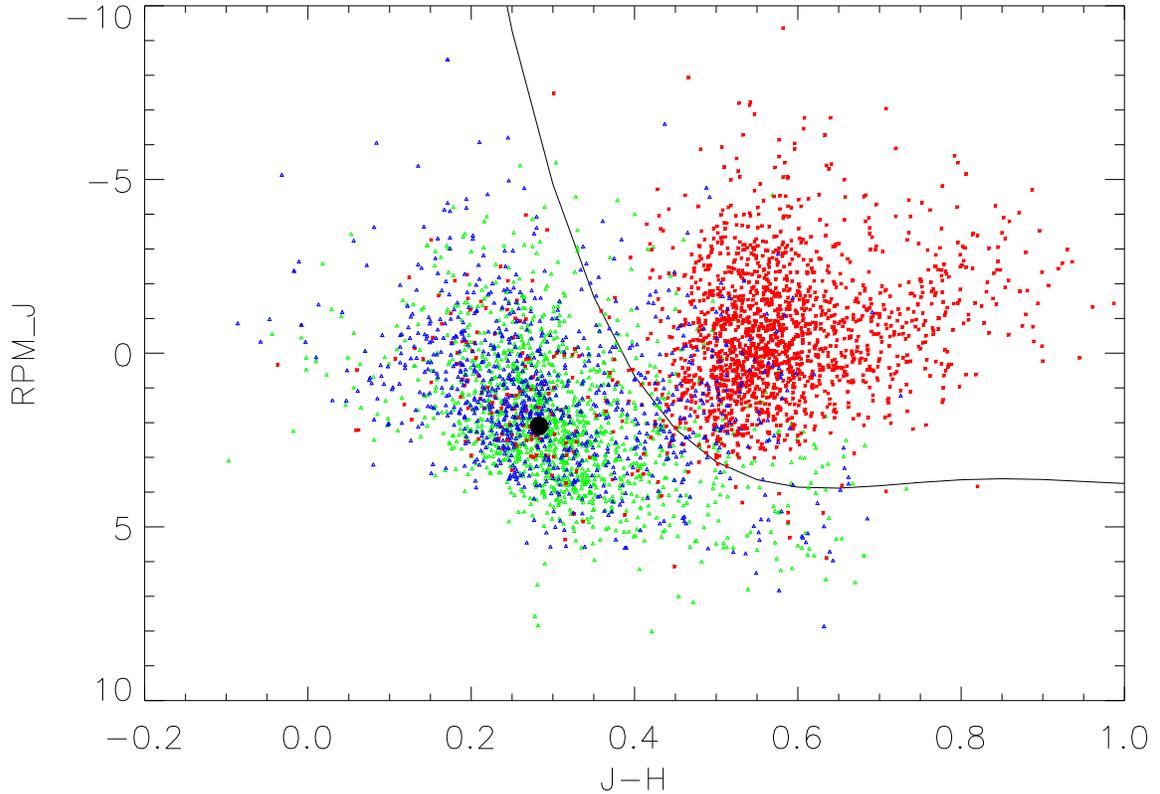}
\caption{\label{fig:rpm}$J$-band reduced proper motion versus $J-H$ color.    Stars from the RAVE DR2 \citep{zwitter08} with galactic latitude, $20^{\circ} \! \leq \! |b| \! < \! 30^{\circ}$, and with measured spectroscopic properties are shown.  The RAVE stars are color coded by luminosity class such that giants ($\log g \! \leq \! 3.5$) are red, dwarfs ($\log g \! > \! 4.1$) are green, and subgiants ($4.1 \! \geq \! \log g \! > \! 3.5$) are blue.   The polynomial relation (solid line) defined from \citet{colliercameron07} discriminates the dwarf star population from the giant star population in this plane.  TYC~1240-00945-1, plotted as the large black circle, is consistent with being a dwarf or subgiant.}
\end{figure}

We can also take advantage of the information provided by the MARVELS
input catalog to place the host star on an RPM
diagram, taking colors from the 2MASS catalog, and proper motions
from the GSC2.3 (see \citealt{gould03} for an example of how RPM can
be used to help differentiate giants from dwarfs).  In Fig.
\ref{fig:rpm}, we show that the $J$-band RPM ($RPM_J \equiv J +
5\log\mu$) is most consistent with the host star being a dwarf or
subgiant, as it falls well away from the region of the RPM
diagram dominated by giant stars.

\subsection{Tidal Effects}
 \label{sec:tides} 

Given the relatively large mass ratio and short period of the TYC
1240-00945-1 system, tidal interactions between the star and MARVELS-1b could be important--  given the roughly $\sim \! 3$\,Gyr age of the host star, is the system likely to be tidally synchronized?  We follow exactly the same analysis of
the tidal interaction as detailed in \citet{fleming10}, which uses the
tidal quality factor of the star, $Q'_*$, as a free parameter in the
equations for the decay of the companion's semimajor axis over time
and the relation of the primary's rotational frequency to the companion's
orbital angular momentum (Eqs.~5 and 6 in \citealt{fleming10});
together, the equations permit a solution for the amount of time
required for tidal synchronization.  Note that if the primary's
rotation never synchronizes, the two bodies may merge
\citep[][]{counselman73,levrard09,jackson09}.  As in
\citet{fleming10}, we have examined the tidal evolution of this system
in the range $10^4 \le Q'_* \le 10^{10}$, for a range of values of the
inclination of the secondary's orbit to the line of sight from $i \! = \! 0^\circ$ (face-on)
to $i \! = \! 90^\circ$ (edge-on), adjusting the mass and rotation period using the measured
values of $v\,\sin\,i$ and $R_*$ from \S\ref{sec:params}.  We set
the primary's equator to be in the same plane as the secondary's
orbit, but this decision does not affect our results.

\begin{figure}
\plotone{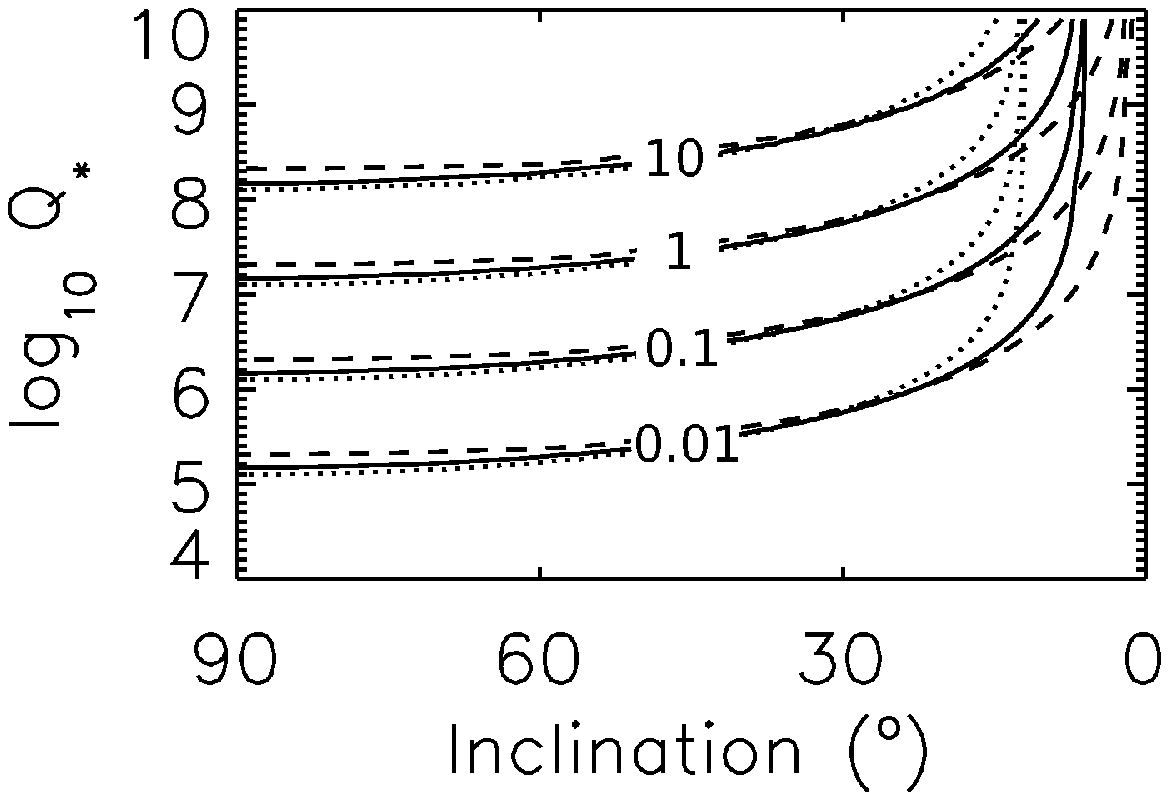}
\caption{\label{fig:tides}Contours of the time (in Gyr) to synchronize the
primary's rotational period to the orbital period, or for the companion
to merge with its host star. Solid curves correspond to the best
fit, dotted curves to the case with $v\,\sin\,i$, $M_*$, and $R_*$ each set at the tops of their $1\sigma$ uncertainty ranges, and dashed curves to the case with $v\,\sin\,i$, $M_*$, and $R_*$ each set at the bottoms of their $1\sigma$ uncertainty ranges.  Merging is only possible in the latter case when $i \! \ge \! 54^\circ$; therefore those portions of the dashed curves represent the
time to merge. }
\end{figure}

In Fig.~\ref{fig:tides} we show the synchronization and merging times from
Eqs.~5 and 6 of \citet{fleming10}, over the $Q'_*$ and $i$ parameter space
defined above.  The curves are isochrones in the ($Q'_*$, $i$) parameter space, so if the TYC 1240-00945-1 system has a ($Q'_*$, $i$) combination that lies above a given isochrone $\tau_{sync/merge}$, then the system will take longer than $\tau_{sync/merge}$ to synchronize or merge.  Isochrones are plotted for $\tau_{sync/merge} = 0.01$, 0.1, 1, and 10 Gyr.

We consider three models:  the best-fit stellar parameters (solid
curves);  one in which $v\,\sin\,i = 3.7$\,km~s$^{-1}$, $M_* = 1.43\,M_\odot$,
and $R_* = 2.44\,R_\odot$ (dotted curves);  and one with $v\,\sin\,i =
0.7$\,km~s$^{-1}$, $M_* = 1.32\,M_\odot$, and $R_* = 2.00\,R_\odot$ (dashed
curves). The latter two cases represent models where the parameter sets were adjusted in opposite directions in an attempt to have the two models span a maximal amount of ($Q'_*, i$) parameter space, while still maintaining the parameters within the uncertainties.  Thus, the uncertainty on the four synchronization/merging isochrones is approximately indicated by the region between the dotted and dashed lines (though it is not a perfect indication of the multi-parameter uncertainty envelope, as is evident from the fact that the dotted and dashed lines cross).

Note if one makes a trial assumption for the value of the inclination
$i$, then given our measurement of $v\,\sin\,i$, one may infer the true
rotational velocity $v$ of the stellar surface.  At some small
inclination, close to a face-on orbit, this will yield a $v$ so
high that the primary's rotational frequency is already spun up to tidal synchronization (and
for the improbable case of an inclination even smaller than this, the primary's rotational frequency is higher than the secondary's orbital frequency, a
scenario we do not explore here, but which would result in gradual spindown of the primary's rotational frequency until it matched with the orbital frequency of the secondary).  For each case of $v\,\sin\,i$ that we
investigated, the value of the inclination which corresponds to
present-day tidal synchronization is visible on Fig.~\ref{fig:tides}
as a vertical asymptote towards which the isochrones converge.  For
inclinations closer to edge-on, the secondary still is in the process
of spinning up the primary.

Next consider the best fit (solid curves) and maximum $v\,\sin\,i$
(dotted curves) cases. We find that for a wide range of low ($Q'_*$, $i$) combinations, the secondary quickly spins the primary up to
synchronization in less time than the $\sim \! 3$\,Gyr age of the host star. However, this alone, while suggestive, is not conclusive proof that such a synchronization has occurred.  As this is an evolved F star, the radius has recently
expanded, complicating any interpretations of the system's
history. Furthermore, for $Q'_* \! \sim \! 10^7$, the synchronization time is
about the age of the system.

For the minimum $v\,\sin\,i$ cases (dashed curves), the rotational
period of the star is very large, $\sim \! 150$ days.  While this period may not
be physical, it is formally permitted by the observations. With such
slow rotation, the companion may merge with the star before synchronization is finished.  This possibility of the synchronization timescale exceeding the merging timescale occurs when $i \! \ge \! 54^\circ$ (note there is no feature in Fig.~\ref{fig:tides} at the $i \! = \! 54^\circ$ transition, because in our simplified model a companion can reach the stellar surface and synchronize the star's rotation period, or move just inside the
surface and merge).  Undoubtedly the behavior of such a compact system
is not well-modeled by Eqs.~5 and 6 of \citet{fleming10} , but we
cannot rule out the possibility that MARVELS-1b will eventually
merge with the host star.

\section{SUMMARY}

In a search through the first year of SDSS-III MARVELS data, we have
discovered MARVELS-1b, a candidate brown dwarf companion to the $V \! \simeq \! 10.6$
star TYC~1240-00945-1 with a velocity semiamplitude of $K=2.533 \pm
0.025$\,km~s$^{-1}$ and an unusually short period of $5.8953 \pm
0.0004$\,d.  Radial velocity data from several observatories
confirm the Doppler variability, and high-resolution spectroscopic
observations indicate that the host is a mildly evolved, slightly
subsolar metallicity F star with $T_{\rm eff} = 6186\pm 92$\,K,
$\log{g}= 3.89 \pm 0.07$, and [Fe/H]=$-0.15 \pm 0.04$, with an inferred
mass of $M_*= 1.37 \pm 0.11\,M_\odot$.  
The minimum mass of MARVELS-1b is $28.0 \pm 1.5\,M_{Jup}$, implying that it is
most likely in the brown dwarf regime.  We see no evidence for
spectral lines from the companion in the high-resolution spectra,
implying that the companion is not an M dwarf with an orbit
extremely close to pole-on. Comprehensive, precise relative photometry
indicates no variability at a level of $\ga \! 1\%$ on time scales of
hours to years.  Phasing to the period of MARVELS-1b as well as the first
harmonic, we can place an upper limit on the amplitude of coherent
photometric variability of $\sim \! 0.05\%$. Under many (but not all) of
the potential combinations of system parameters, this short-period
system is likely to have tidally synchronized, given the estimated
$\sim \! 3$\,Gyr age of the host star.

The {\it a priori} transit probability of MARVELS-1b is quite high, $\sim \! 14\%$.  Although we find no evidence for transits, we also cannot definitively rule them out for likely MARVELS-1b radii of $r \! \sim \! R_{Jup}$.  The transit ephemeris is $T_C=2454936.555 \pm 0.024$\,(BJD$_{\rm TDB}$), with an expected transit depth of $\sim \! 0.2\% (r/R_{Jup})^2$, and a duration of $\sim \! 6.5$\,hours for a central transit.

We believe this candidate highlights the great promise of MARVELS as
a factory for finding the rare companions that populate the brown
dwarf desert.  The primary goal of the MARVELS survey is to monitor
$\sim \! 10^4$ main sequence and subgiant stars with velocity precision
sufficient to detect Jovian companions with periods of less than a few
years.  As such, MARVELS is uniquely and exquisitely sensitive to
massive but rare companions.  MARVELS-1b is the first of a
number of brown dwarf candidates we have identified in the MARVELS
data obtained to date, and we expect to uncover several additional
such systems as the survey progresses.

\begin{table}[htbp]
\begin{center}
\caption{{ SDSS-III MARVELS Radial Velocities for TYC~1240-00945-1 \label{tab:obsjournal}}}
\begin{tabular}{lcccccc}
\hline\hline
BJD$_{\rm TDB}$ & Differential & Stat. err. & Scaled err. & Differential & Stat. err. & Scaled err. \\
 & RV$_{\rm beam1}$ (km~s$^{-1}$)  & (km~s$^{-1}$)  & (km~s$^{-1}$) & RV$_{\rm beam2}$ (km~s$^{-1}$)  &  (km~s$^{-1}$)  &  (km~s$^{-1}$)  \\
\hline
2454777.81083 & -1.15 & 0.05 & 0.11 & -1.16 & 0.05 & 0.18\\ 
2454778.78470 & -2.81 & 0.04 & 0.10 & -2.89 & 0.04 & 0.16\\ 
2454779.74062 & -1.52 & 0.03 & 0.07 & -1.49 & 0.03 & 0.12\\ 
2454781.65432 & 2.37 & 0.05 & 0.11 & 2.23 & 0.05 & 0.18\\ 
2454785.83590 & -0.94 & 0.04 & 0.09 & -0.96 & 0.04 & 0.15\\ 
2454786.88843 & 1.57 & 0.04 & 0.10 & 1.48 & 0.05 & 0.16\\ 
2454787.85523 & 2.42 & 0.04 & 0.09 & 2.35 & 0.04 & 0.15\\ 
2454787.90098 & 2.38 & 0.06 & 0.14 & 2.38 & 0.06 & 0.23\\ 
2454840.69407 & 2.49 & 0.04 & 0.08 & 2.60 & 0.04 & 0.13\\ 
2454841.67278 & 1.20 & 0.04 & 0.09 & 1.36 & 0.04 & 0.14\\ 
2454842.65535 & -1.13 & 0.05 & 0.10 & -1.13 & 0.05 & 0.17\\ 
2454843.68547 & -2.83 & 0.05 & 0.12 & -2.80 & 0.05 & 0.20\\ 
2454844.69581 & -1.23 & 0.04 & 0.10 & -1.14 & 0.04 & 0.16\\ 
2454868.61695 & -0.26 & 0.04 & 0.08 & -0.27 & 0.04 & 0.13\\ 
2454869.60690 & 1.99 & 0.04 & 0.10 & 1.97 & 0.04 & 0.16\\ 
2455141.74609 & 2.16 & 0.03 & 0.06 & 2.05 & 0.03 & 0.10\\ 
2455142.78463 & 0.06 & 0.04 & 0.09 & 0.04 & 0.04 & 0.15\\ 
2455143.76503 & -2.22 & 0.03 & 0.07 & -2.10 & 0.03 & 0.11\\ 
2455144.80421 & -2.36 & 0.03 & 0.06 & -2.28 & 0.03 & 0.10\\ 
2455145.80876 & -0.20 & 0.04 & 0.08 & -0.23 & 0.04 & 0.13\\ 
\hline
\end{tabular}
\end{center}
\end{table}

\begin{table}[htbp]
\begin{center}
\caption{{ TYC~1240-00945-1:  Parameters of the Star \label{tab:stellarparams}}}
\begin{tabular}{lc}
\hline\hline
Parameter & Value \\
\hline
Spectral Type & F9IV-V \\
$g'$ & 10.821 $\pm$ 0.013 \\
$r'$ & 10.436 $\pm$ 0.007 \\
$i'$ & 10.324 $\pm$ 0.013 \\
$B$ & 11.230 $\pm$ 0.025 \\
$V$  & 10.612 $\pm$ 0.025 \\
$R_c$ & 10.242 $\pm$ 0.011 \tablenotemark{a} \\
$I_c$ & 9.916 $\pm$ 0.011 \tablenotemark{a} \\
$J_{2MASS}$ & 9.395 $\pm$ 0.018 \\
$H_{2MASS}$ & 9.112 $\pm$ 0.016 \\
$K_{2MASS}$ & 9.032 $\pm$ 0.017 \\
$T_{\rm eff}$ & 6186 $\pm$ 92 K\\
$\log g$ & 3.89 $\pm$ 0.07 (cgs)\\
$\rm [Fe/H]$ & -0.15 $\pm$ 0.04\\
Mass & 1.37 $\pm$ 0.11 $M_{\odot}$\\ 
Radius & $2.20^{+0.25}_{-0.22} R_{\odot}$\\ 
$A_V$ & $0.40 \pm 0.05$\\
Distance & $280 \pm 30$ pc\\
$v\,\sin\,i$ & $2.2 \pm 1.5$\,km~s$^{-1}$\\
\hline
\tablenotetext{a}{$R_cI_c$ are transformed magnitudes based on $g'r'i'$, using the transformation equations of \citet{smith02}.}
\end{tabular}
\end{center}
\end{table}

\begin{table}[htbp]
\begin{center}
\caption{{ HET Radial Velocities for TYC~1240-00945-1 \label{tab:obsjournalhet}}}
\begin{tabular}{lccc}
\hline\hline
BJD$_{\rm TDB}$ & Differential RV & Stat. error & Scaled error \\
 & (km~s$^{-1}$)  & (km~s$^{-1}$)  & (km~s$^{-1}$)  \\
\hline
2455175.59389 & 0.78 & 0.02 & 0.29\\ 
2455177.61679 & 1.01 & 0.02 & 0.25\\ 
2455178.60575 & -1.26 & 0.02 & 0.25\\ 
2455180.80563 & -0.83 & 0.02 & 0.25\\ 
2455181.79358 & 1.31 & 0.01 & 0.23\\ 
2455182.79353 & 2.50 & 0.02 & 0.38\\ 
2455183.58448 & 0.74 & 0.02 & 0.24\\ 
2455184.58343 & -1.49 & 0.02 & 0.26\\ 
2455185.57637 & -2.72 & 0.02 & 0.34\\ 
\hline
\end{tabular}
\end{center}
\end{table}

\begin{table}[htbp]
\begin{center}
\caption{{ SMARTS Absolute Radial Velocities for TYC~1240-00945-1 \label{tab:obsjournalsmarts}}}
\begin{tabular}{lccc}
\hline\hline
BJD$_{\rm TDB}$ & Absolute RV & Stat. error & Scaled error \\
 & (km~s$^{-1}$)  & (km~s$^{-1}$)  & (km~s$^{-1}$)  \\
\hline
2455052.89667 & 19.7 & 0.2 & 0.3\\ 
2455053.91287 & 18.9 & 0.3 & 0.4\\ 
2455084.78037 & 16.6 & 0.2 & 0.3\\ 
2455093.77977 & 19.9 & 0.2 & 0.3\\ 
2455109.72687 & 16.3 & 0.2 & 0.3\\ 
2455112.80387 & 18.8 & 0.3 & 0.5\\ 
2455139.64937 & 16.4 & 0.4 & 0.6\\ 
2455140.75867 & 19.0 & 0.3 & 0.4\\ 
2455164.71377 & 19.9 & 0.2 & 0.3\\ 
\hline
\end{tabular}
\end{center}
\end{table}

\begin{table}[htbp]
\begin{center}
\caption{{ MARVELS-1b:  Parameters of the Companion \label{tab:companionparams}}}
\begin{tabular}{cc}
\hline\hline
Parameter & Value \\
\hline
Minimum Mass & 28.0 $\pm$ 1.5 $M_{Jup}$\\
$a$ & 0.071 $\pm$ 0.002 AU \\
$K$ & 2.533 $\pm$ 0.025 km~s$^{-1}$ \\
$P$ & 5.8953 $\pm$ 0.0004 d \\
$T_{\rm{prediction\; for\; transit}}$ & 2454936.555 $\pm$ 0.024 (BJD$_{\rm TDB}$)\\
$e\cos\omega$ & -0.015 $^{+0.010}_{-0.010}$ \\
$e\sin\omega$ & -0.003 $^{+0.008}_{-0.009}$ \\
\hline
\end{tabular}
\end{center}
\end{table}

\begin{table}[htbp]
\begin{center}
\caption{{ TYC~1240-00945-1 Individual Determinations of Stellar Parameters\label{tab:stellarparamsraw}}}
\begin{tabular}{cccccl}
\hline\hline
$T_{\rm eff}$ & $\log g$ & [Fe/H] & $\xi_{t}$ & $v\sin i$ & Notes \\
(K) & (cgs) &  & (km~s$^{-1}$) & (km~s$^{-1}$) &  \\
\hline
6186 $\pm$ 82 & 4.01 $\pm$ 0.17 & -0.14 $\pm$ 0.08 & 1.26 $\pm$ 0.17 & 2.2 $\pm$ 1.5 & High-res. (ESO 2.2-m)\\
6090 $\pm$ 74 & 3.89 $\pm$ 0.13 & -0.21 $\pm$ 0.06 & 1.13 $\pm$ 0.18 & - & High-res. (APO 3.5-m, hand redux)\\
6274 $\pm$ 112 & 3.89 $\pm$ 0.22 & -0.15 $\pm$ 0.09 & - & $\lesssim 9$ & High-res. (APO 3.5-m, SME redux)\\
$6400^{+400}_{-600}$ & 3.5 $\pm$ 1.5 & 0.0 $\pm$ 2.0 & - & - & SED fit to photometry\\
\hline
\end{tabular}
\end{center}
\end{table}

\acknowledgments

Funding for the MARVELS multi-object Doppler instrument was provided by the W.M. Keck Foundation and NSF with grant AST-0705139.
The MARVELS survey was partially funded by the SDSS-III consortium, NSF grant AST-0705139, NASA with grant NNX07AP14G and the University of Florida. 
Funding for SDSS-III has been provided by the Alfred P. Sloan Foundation, the Participating Institutions, the National Science Foundation, and the U.S. Department of Energy. The SDSS-III web site is \url{http://www.sdss3.org/}.
SDSS-III is managed by the Astrophysical Research Consortium for the Participating Institutions of the SDSS-III Collaboration including the University of Arizona, the Brazilian Participation Group, University of Cambridge, University of Florida, the French Participation Group, the German Participation Group, the Michigan State/Notre Dame/JINA Participation Group, Johns Hopkins University, Lawrence Berkeley National Laboratory, Max Planck Institute for Astrophysics, New Mexico State University, New York University, the Ohio State University, University of Portsmouth, Princeton University, University of Tokyo, the University of Utah, Vanderbilt University, University of Virginia, University of Washington and Yale University.
The Hobby-Eberly Telescope (HET) is a joint project of the University
of Texas at Austin, the Pennsylvania State University, Stanford
University, Ludwig-Maximilians-Universit\"{a}t M\"{u}nchen, and
Georg-August-Universit\"{a}t G\"{o}ttingen. The HET is named in honor of its
principal benefactors, William P. Hobby and Robert E. Eberly.
The authors thank Debra Fischer for kindly providing a preliminary
version of her precise Doppler pipeline for use with HRS.
FEROS spectra were observed at the ESO 2.2 m telescope under the ESO-ON agreement.
This research is partially supported by funding from the Center for Exoplanets
and Habitable Worlds.  
The Center for Exoplanets and Habitable Worlds is supported by the
Pennsylvania State University, the Eberly College of Science, and the
Pennsylvania Space Grant Consortium.
Keivan Stassun, Leslie Hebb, and Joshua Pepper acknowledge funding support from the Vanderbilt Initiative in Data-Intensive Astrophysics (VIDA) from Vanderbilt University, and from NSF Career award AST-0349075.
EA thanks NSF for CAREER grant 0645416.
GFPM acknowledges financial support from CNPq grant n$^{\circ}$ 476909/2006-6 and FAPERJ grant n$^{\circ}$ APQ1/26/170.687/2004.
J.P.W. acknowledges support from NSF Astronomy \& Astrophysics Postdoctoral Fellowship AST 08-02230.



\end{document}